\documentclass[twocolumn,prd]{revtex4-2}
\usepackage[english]{babel}
\usepackage{slashed}

\usepackage[letterpaper,top=2cm,bottom=2cm,left=3cm,right=3cm,marginparwidth=1.75cm]{geometry}

\usepackage{amsmath}
\usepackage{graphicx}
\usepackage[colorlinks=true, allcolors=blue]{hyperref}
\newcommand{\gev}{\mathrm{GeV}}
\newcommand{\met}{E\!\!\!/_T}
\begin{document}
\title{Study of the $tH$ production at 14TeV and 100TeV $pp$ colliders}
\author{Yi-Dong Song}
\email{songyd@mail.sdu.edu.cn}
\affiliation{School of Physics, Shandong University, Jinan, 250100, P. R. China}
\author{Shou-Shan Bao}
\email{ssbao@sdu.edu.cn}
\affiliation{Institute of Frontier and Interdisciplinary Science, Shandong University, Qingdao, 266237, P. R. China}
\begin{abstract}
The flavor-changing neutral current interactions in the standard model are suppressed seriously and such interactions can be used to search the new physics beyond SM. The top quark and Higgs bosons are heavier than the other particles in SM, we can expect the new physics plays a more important role in their interactions. In this work, we study the flavor- changing neutral current interactions between the top quark $\bar{t}q H$ through the production of the single top associated with a Higgs boson on 14TeV and 100TeV $pp$ colliders. We consider the leptonic decay channels of Higgs and study the signal. We find a sensitive region $0.4\leq\Delta R\leq1.4$ for the leptons from Higgs bosons and $\Delta R\geq1.8$ between the jets from top quark and leptons. We investigate the detective abilities of the hadron colliders for the processes. Although this process seems not comparable with $pp\to t\bar{t}$ with $t\to H q$ decay, it is still attractive since this process can be used to distinguish the $y_{tu}$ and $y_{tc}$. 
\end{abstract}
\maketitle

\section{Introduction}
The discovery of the Higgs boson at the Large Hadron Collider (LHC)\cite{Aad:2012tfa,Chatrchyan:2012xdj} is a great triumph of the Standard Model (SM). Following that, one can expect to test and search for new physics through the precision measurement of the Higgs boson properties. In SM, the Higgs doublet is introduced to generate masses for the gauge bosons and the fermions. Most parameters in SM are related to Yukawa couplings. Since the top quark and Higgs boson are heavier than the other particles in SM, we can expect that high energy new physics may play more important roles in their couplings and the Yukawa coupling between the top quark and Higgs boson could provide better clues to study the mechanism of electroweak symmetry breaking and new physics beyond SM. On the other hand, so far there is no signal of new physics particles has been found in direct searches. Therefore indirect search through the precision measurement can be expected to constrain the new physics scale or coupling. In SM the flavor-changing neutral current (FCNC) couplings are absent at the tree level and suppressed at the loop level. So that such processes are thought to be good clues to search for the new physics beyond SM indirectly.

The FCNC interactions of top quark are small and the branch ratios of such decay are predicted as $Br(t\to Hc)\sim 10^{-14}$ and $Br(t\to Hu)\sim 10^{-17}$ in SM\cite{Mele:1998ag, Eilam:1990zc}, which are far below the sensitivity of LHC. Many new physics extensions to the SM can enhance such interaction at tree level or loop level, e.g. two Higgs doublet model\cite{Wu:1994ja,Atwood:1996vj,Bao:2008hd}, supersymmetry model\cite{Haber:1984rc, Cao:2014udj, Dedes:2014asa} and other new physics models\cite{AguilarSaavedra:2004wm}. One can expect that the new physics contributions would play significant roles in such FCNC interactions and such contributions could be discovered or constrained on high energy colliders.


%


In this work, we focus on the effective FCNH interaction between the Higgs boson and the top quark which can be expressed in general as~\cite{Chen:2013qta,Kao:2011aa}
\begin{equation}
-\mathcal{L}_Y\supset\sum_{q=u,c}y_{tq}\bar{q} t H + h.c., \label{eq:lgr}
\end{equation}
where the $y_{tq}$ is the FCNC Yukawa coupling constant. Such anomalous couplings have been studied widely. Many works have been done through $t\to Hq$ rare decays~\cite{Kao:2011aa,Atwood:2013ica,Han:2001ap}, the single top quark production with Higgs boson~\cite{Wu:2014dba,Wang:2012gp,Greljo:2014dka}, and same sign top quark production~\cite{Atwood:2013xg}. Recently, a study invoving $pp\to t H$ and $pp\to tqH$ on 100TeV hadron collider is done in Ref.~\cite{Ozsimsek:2022lte}.
Through $pp\to t\bar{t}$ with $t\to H q$ channel, ATLAS\cite{ATLAS:2022qxx} and CMS\cite{CMS:2021gfa} set upper limits at the 95\% confidence level (C.L.) on the couplings as
\begin{align}
\text{ATLAS:~}& Br(t\to H c)<9.9\times10^{-4},\\
						  & Br(t\to H u)<7.2\times10^{-4},\\
\rm{CMS:~}    & Br(t\to H c)<9.4\times10^{-4},\\
							& Br(t\to H u)<7.9\times10^{-4}.
\end{align}
Since the $u$-quark and $c$-quark from top quark decay can be distinguished, the LHC can not distinguish the vertexes $\bar{t}u H$ and $\bar{t}c H$. However, the two vertexes can be distinguished in $tH$ and $\bar{t}H$ productions, since the PDF luminosity of the $u$-quark in the proton is larger than the sea quark. 

In this work we focus on the process $pp\to t/\bar{t} H$ at high luminosity LHC and 100TeV $pp$ colliders in future~\cite{CEPCStudyGroup:2018ghi,CEPCStudyGroup:2018ghi}. In Ref.\cite{Wang:2012gp}, the QCD NLO corrections have been studied at LHC. In Ref.~\cite{Bao:2019hor} this process was analyzed with $H\to b\bar{b}$ channel, and it was found that the angular of the final states could be used to separate the boosted Higgs signal from the main background. Since the isolated high $p_t$ electrons and muons are typically clean and easy detected on colliders, we focus on the $h\to VV^*$ ($V=W, Z$) channels to study the $tH$ associated production at hadron colliders and try to examine whether such channels are prossiming channels for measurement of the anomalous Yukawa couplings at $pp$ colliders. 

This work is organized as follows. In Sec. II, the $H\to WW^*\to \ell^+\ell^-+\met$ channel is analyzed for 14TeV and 100TeV $pp$ colliders. In Sec. III, the $H\to ZZ^*\to 4\ell^\pm$ channel is argued. The results and the discussion are given in Sec. IV and a short summary is given in Sec. V.
\section{$h\to WW^*\to 2\ell^\pm+\met$}
\begin{figure}%
\centering
\includegraphics[width=\columnwidth]{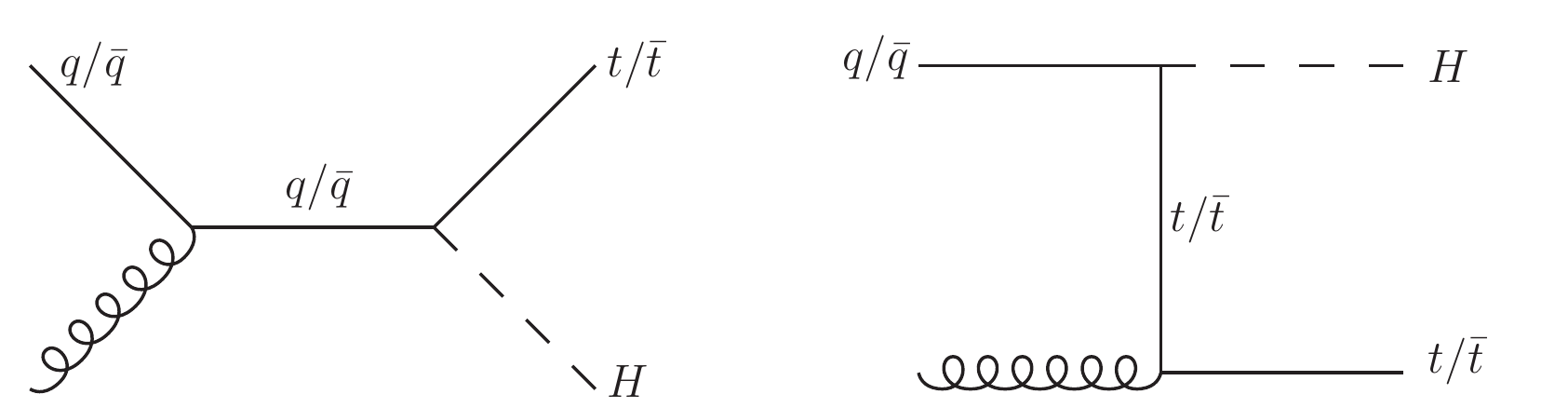}%
\caption{The Feynman diagram for the $tH$ and $\bar{t}H$ production at hadron colliders.}%
\label{fg:fmd}%
\end{figure}

We study the anomalous interaction in Eq.~\eqref{eq:lgr} through $pp\to t/\bar{t}H$ and the Feynman diagrams are shown in Fig.~\ref{fg:fmd}. Such a process can also be induced by the FCNC coupling of strong interaction $\bar{t}q g$ in some new physics models. However, there will be only $s$-channel contribution which is suppressed at high energy colliders. 

In this work, we focus on the leptonic decay channels of Higgs and hadronic decays of the top quark. Firstly we consider the $h\to WW^*$ channel. The final states are characterized by two charged leptons and three jets including one $b$-jet and two light jets. 
The dominate background is the process $pp\to t \bar{t}j\to WWb\bar{b}j$ which is denoted as $t\bar{t}$ in this work. Also the processes of EW gauge bosons produced with light jets $W^+ W^- +3j$ and $pp\to Z+3j$ are considered, which can be suppressed by the b-tag and the invariant mass of the two charged leptons. 

We generate the events of the signals and backgrounds with Madgraph~\cite{Alwall:2014hca} with Pythia8~\cite{Bierlich:2022pfr} and Deltphes3~\cite{deFavereau:2013fsa}. The default values in Madgraph for the SM parameters are used. The build-in ATLAS card in Delphes is used for the simulation of 14 TeV hadron collider and the FCC-h card is used for 100TeV hadron collider. We choose NNPDF23LO~\cite{Ball:2012cx} parton distribution function set in this work.

In the analysis we need $N_j\geq3, N_l\geq2$ and $N_b=1$, and some other basic acceptance cuts are needed for the selection of events,
\begin{subequations}\label{eq:WW_I}
\begin{align}
&P_T\geq25 \gev,\quad E\!\!\!/_T>25 \gev,\\
&\Delta R\geq0.4,\quad \lvert\eta\rvert\leq2.5,
\end{align}
\end{subequations}
where the $p_T$ and $\eta$ are the transverse momentum and rapidity of leptons and jets. $E\!\!\!/_T$ is the missing transverse energy. The angular distance $\Delta R=\sqrt{(\Delta \eta)^2+(\Delta \phi)^2}$ is the separation in $\eta-\phi$ plane between different final states. Indeed, the acceptance cuts depend on the detector could be different for the jets and leptons. Here, we employ the same cuts for different final states for simplicity. All the above cuts are denoted as Cut-I in our analysis.

\begin{figure*}[htb]
    \centering 
    \includegraphics[width=0.3\textwidth,trim={25 5 50 35},clip]{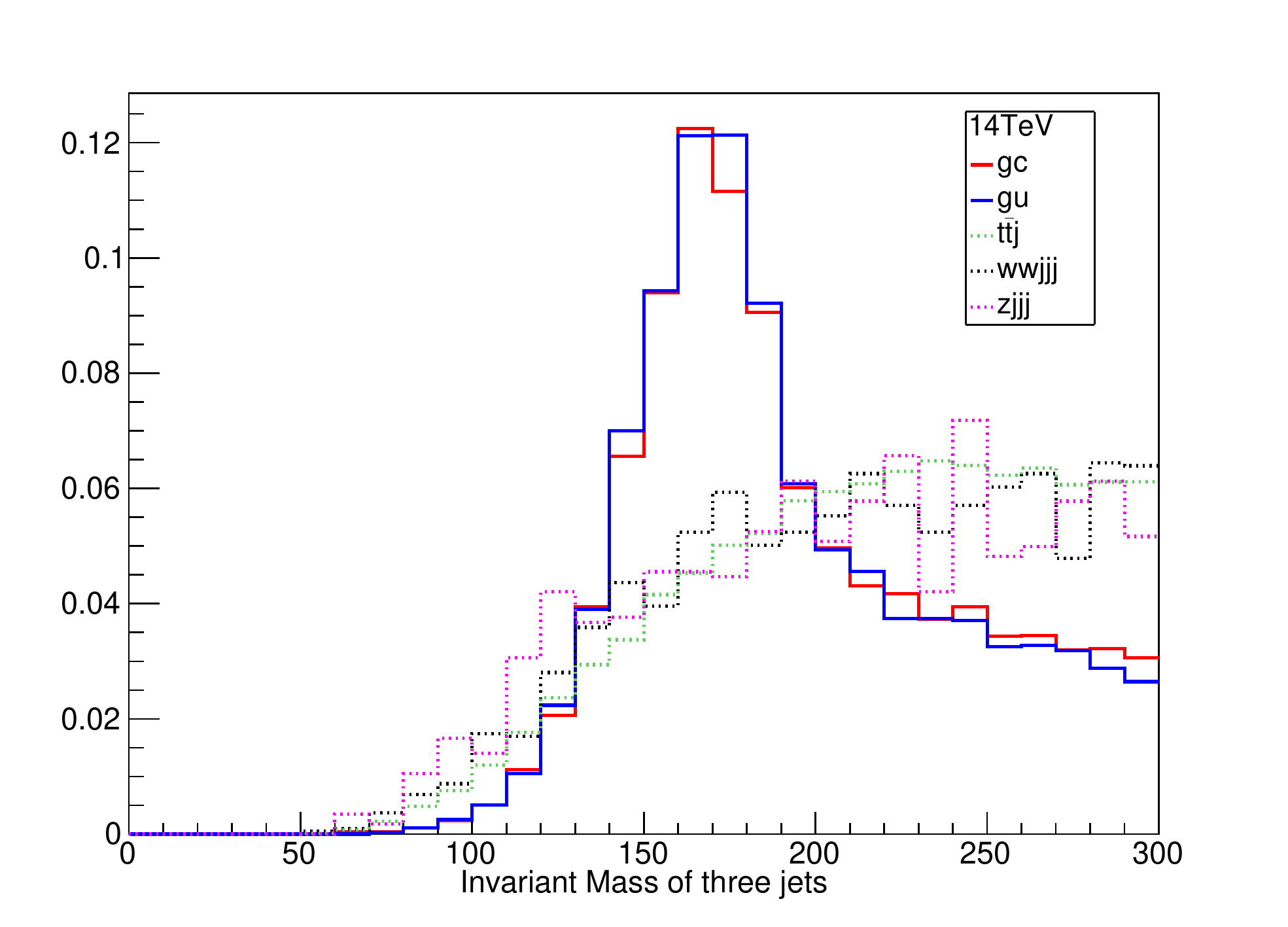}
    \includegraphics[width=0.3\textwidth,trim={25 5 50 35},clip]{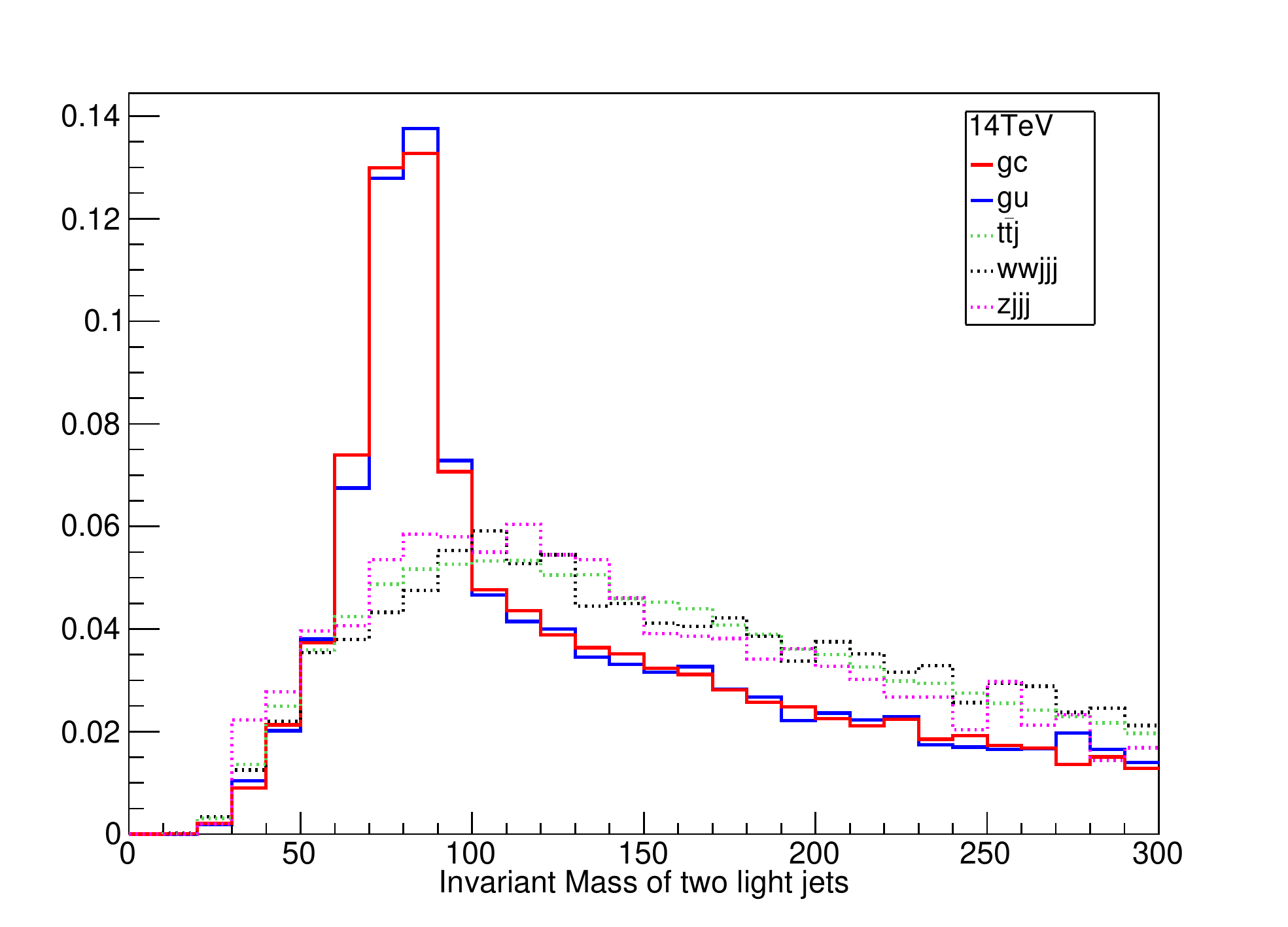}
    \includegraphics[width=0.3\textwidth,trim={25 5 50 35},clip]{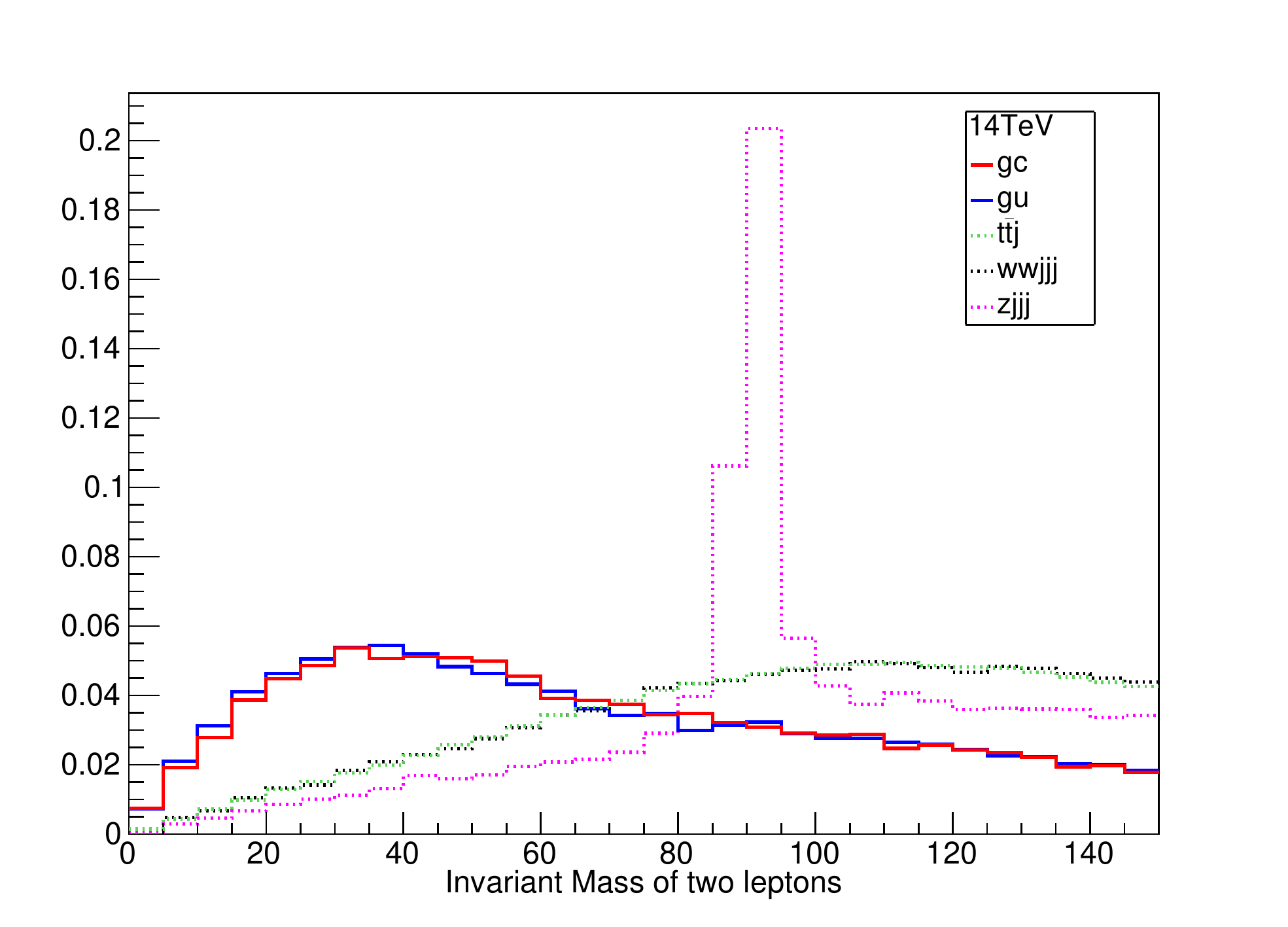}\\
    \includegraphics[width=0.3\textwidth,trim={25 5 50 35},clip]{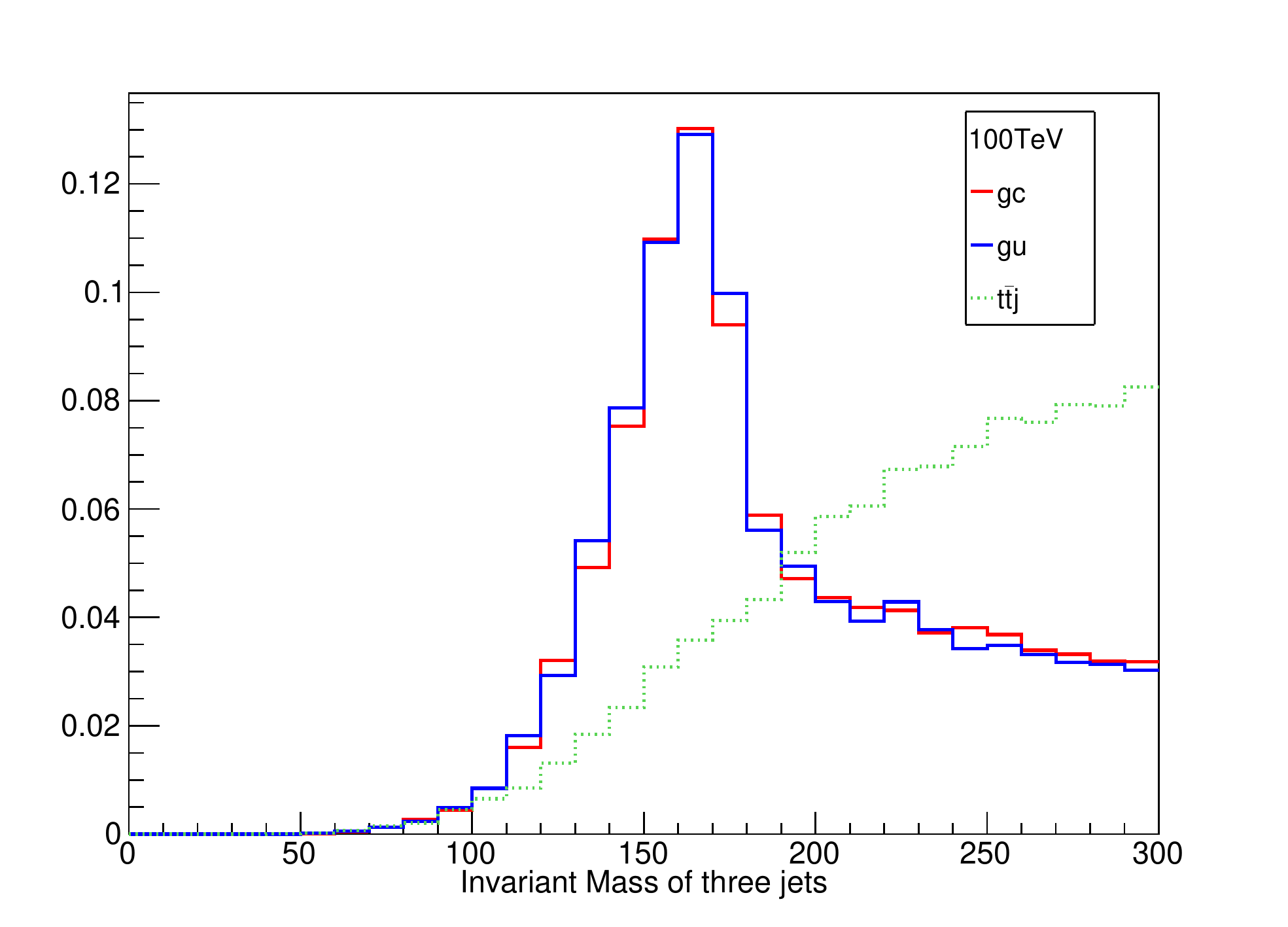}
    \includegraphics[width=0.3\textwidth,trim={25 5 50 35},clip]{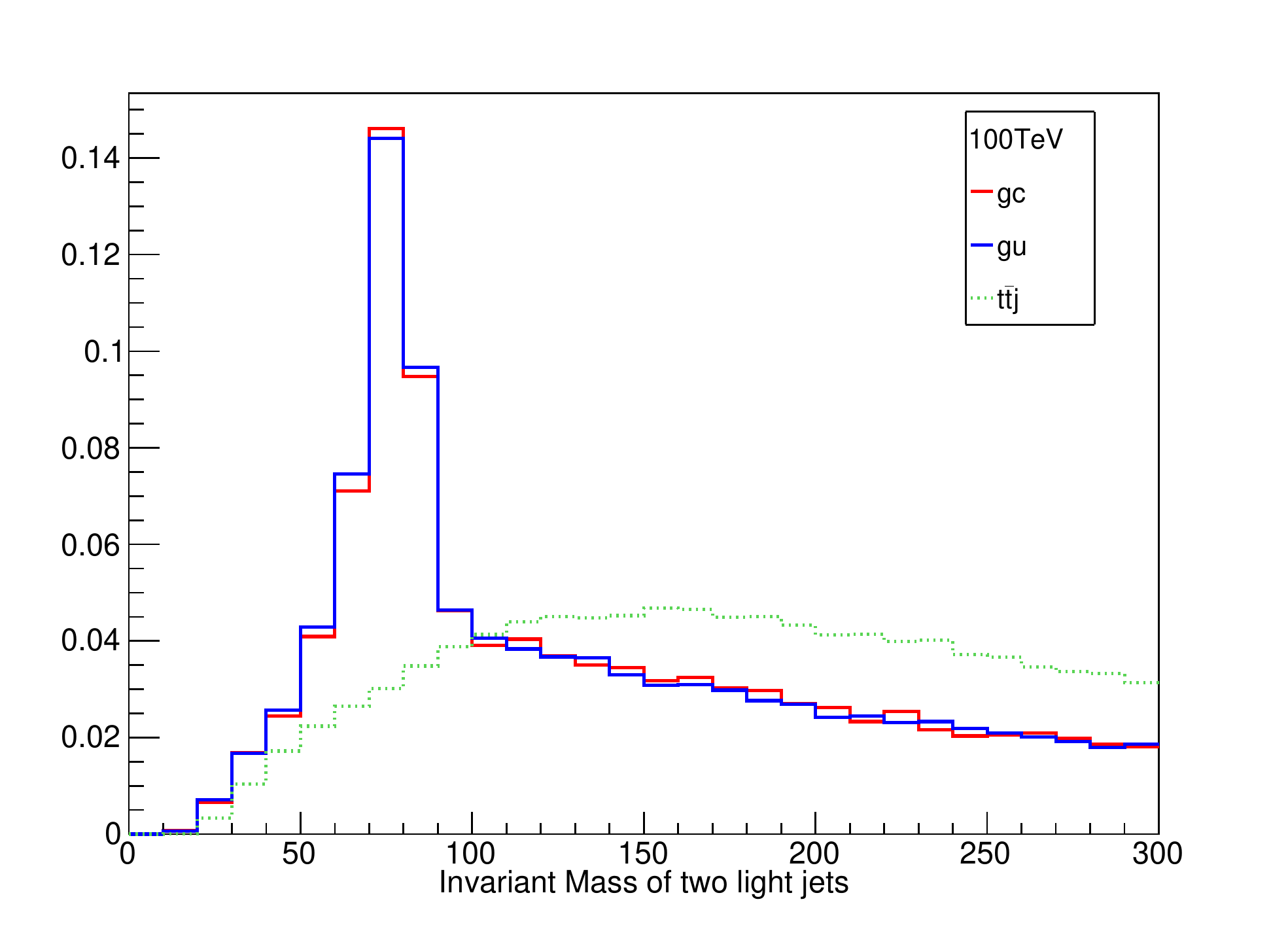}
    \includegraphics[width=0.3\textwidth,trim={25 5 50 35},clip]{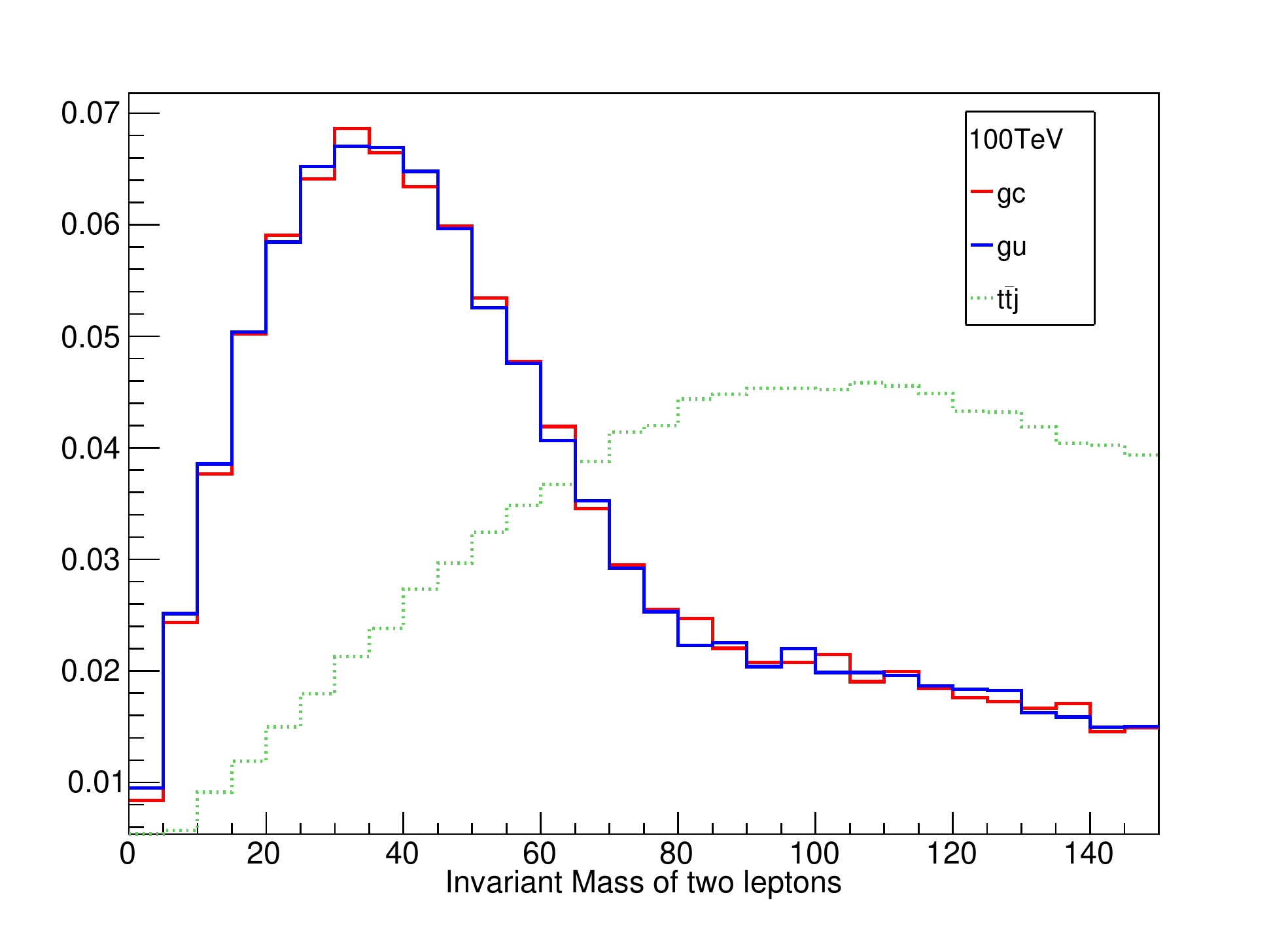}
\caption{The normalized distributions of the invariant mass (GeV) of the three jets, two light jets and two charged leptons at 14TeV and 100 TeV $pp$ colliders.}%
\label{fg:invariantmass}%
\end{figure*}

To reconstruct the top quark and Higgs boson, we need other selection conditions. 
In Fig.~\ref{fg:invariantmass}, the distributions for the invariant mass of the three jets and that of two charged leptons are shown.
Since the $b$-jet and two light jets are from one top quark, we can expect the invariant mass $m_{bjj}$ to be close to the top mass. Furthermore, the two light jets are from one $W$ boson, so we can employ a cut on the invariant mass $m_{jj}$ near $M_W$. The three jets in the $t\bar{t}j$ background are from two different top quarks where one of the two $b$-jets is miss tagged. Since then there is no peak in the distributions of the backgrounds. According to the plots, we can apply the cuts for the $m_{bjj}$ and the $m_{jj}$ to reconstruct the top quark as
\begin{subequations}\label{eq:wwtop}
\begin{align}
130\gev\leq m_{bjj}\leq200 \gev.\\
50\gev\leq m_{jj}\leq100 \gev.
\end{align}
\end{subequations}
The two charged leptons which are from the Higgs boson, carry a part of the Higgs energy while the other part is carried by the neutrinos. Since then there will be $m_{\ell\ell}<M_W$. However, there will be a narrow peak near $M_Z$ in the $m_{\ell\ell}$ distribution for the background $Zjjj$. It's easy to veto such events with an up limit cut on the $m_{\ell\ell}$ as
\begin{align}
m_{\ell\ell}\leq81\gev. \label{eq:wwhiggs}
\end{align} 
We categorize the cuts of $m_{bjj}$, $m_{jj}$ and $m_{\ell\ell}$ as Cut-II. 

\begin{figure*}[htb]%
    \centering 
    \includegraphics[width=0.3\textwidth,trim={25 10 50 40},clip]{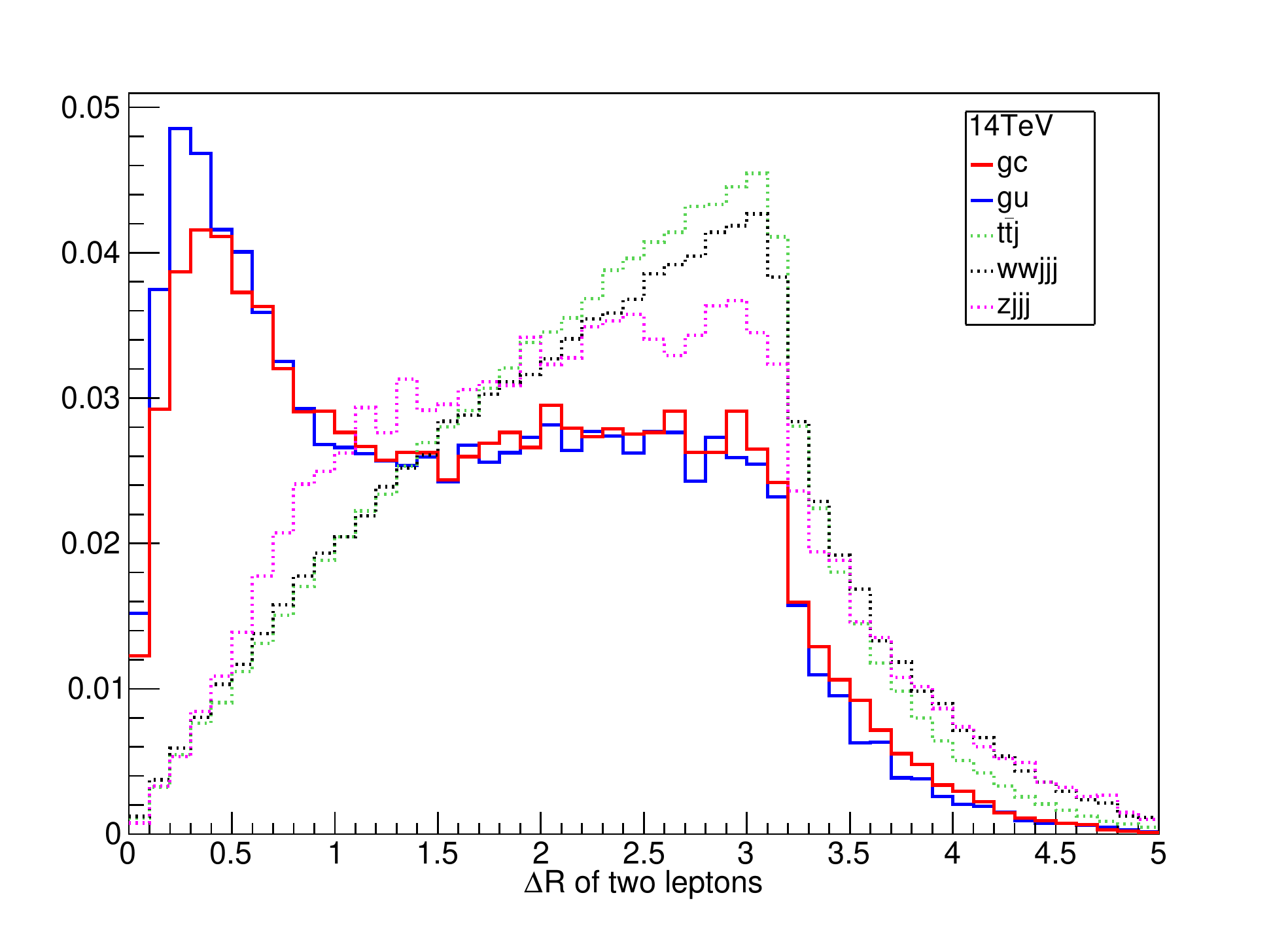}
    \includegraphics[width=0.3\textwidth,trim={25 10 50 40},clip]{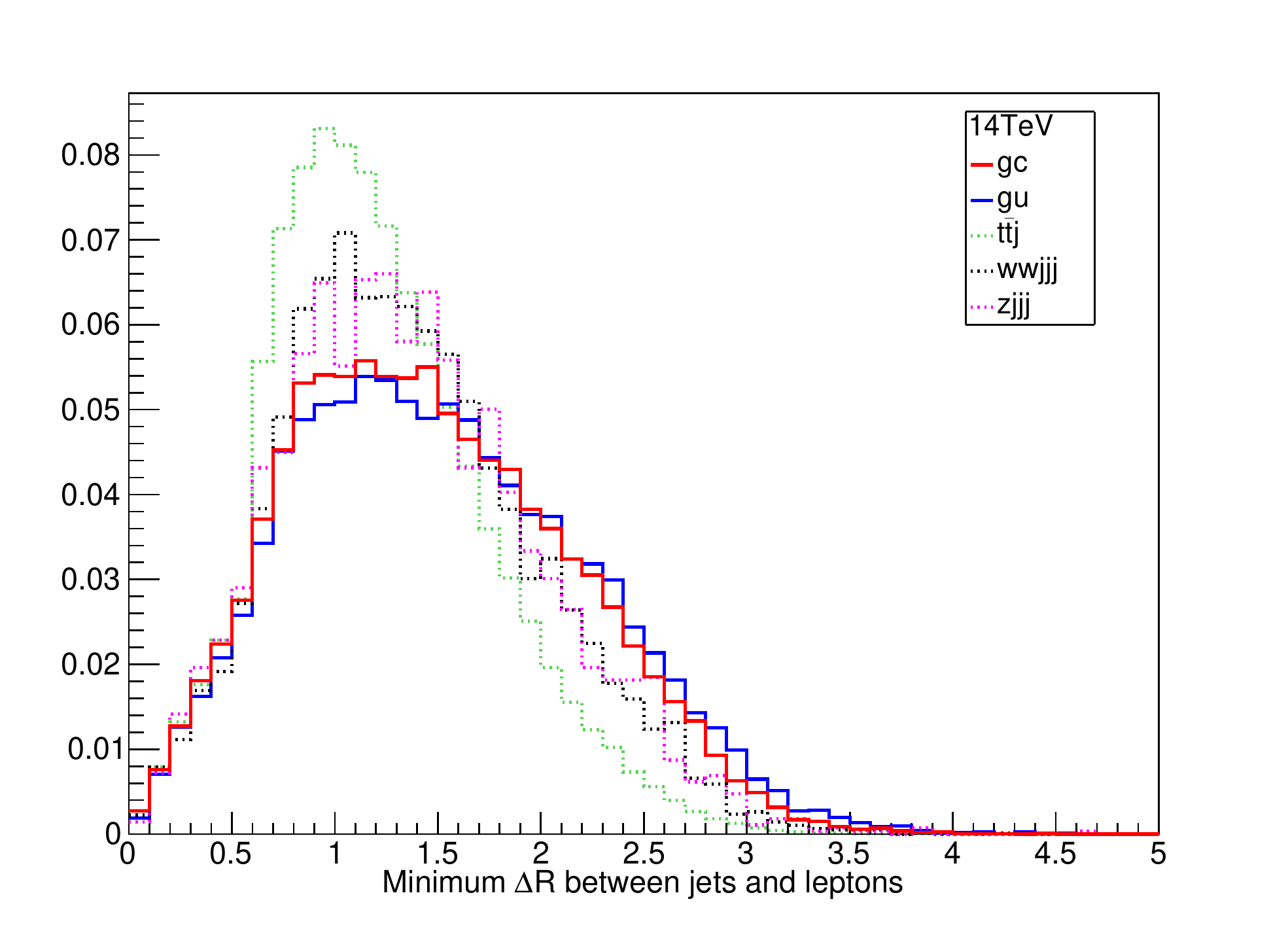}\\
		\includegraphics[width=0.3\textwidth,trim={22 10 53 40},clip]{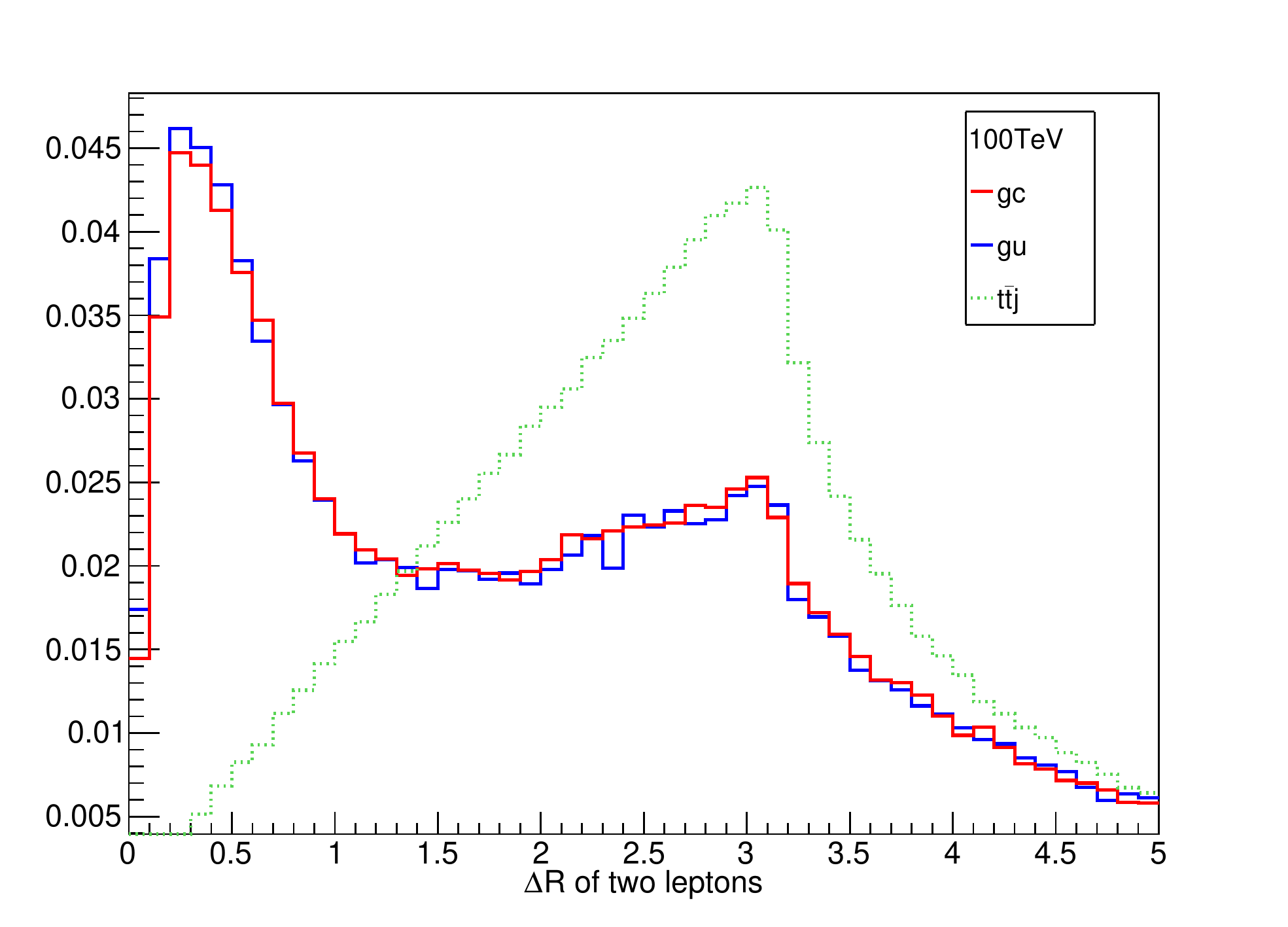} 
    \includegraphics[width=0.3\textwidth,trim={25 10 50 38},clip]{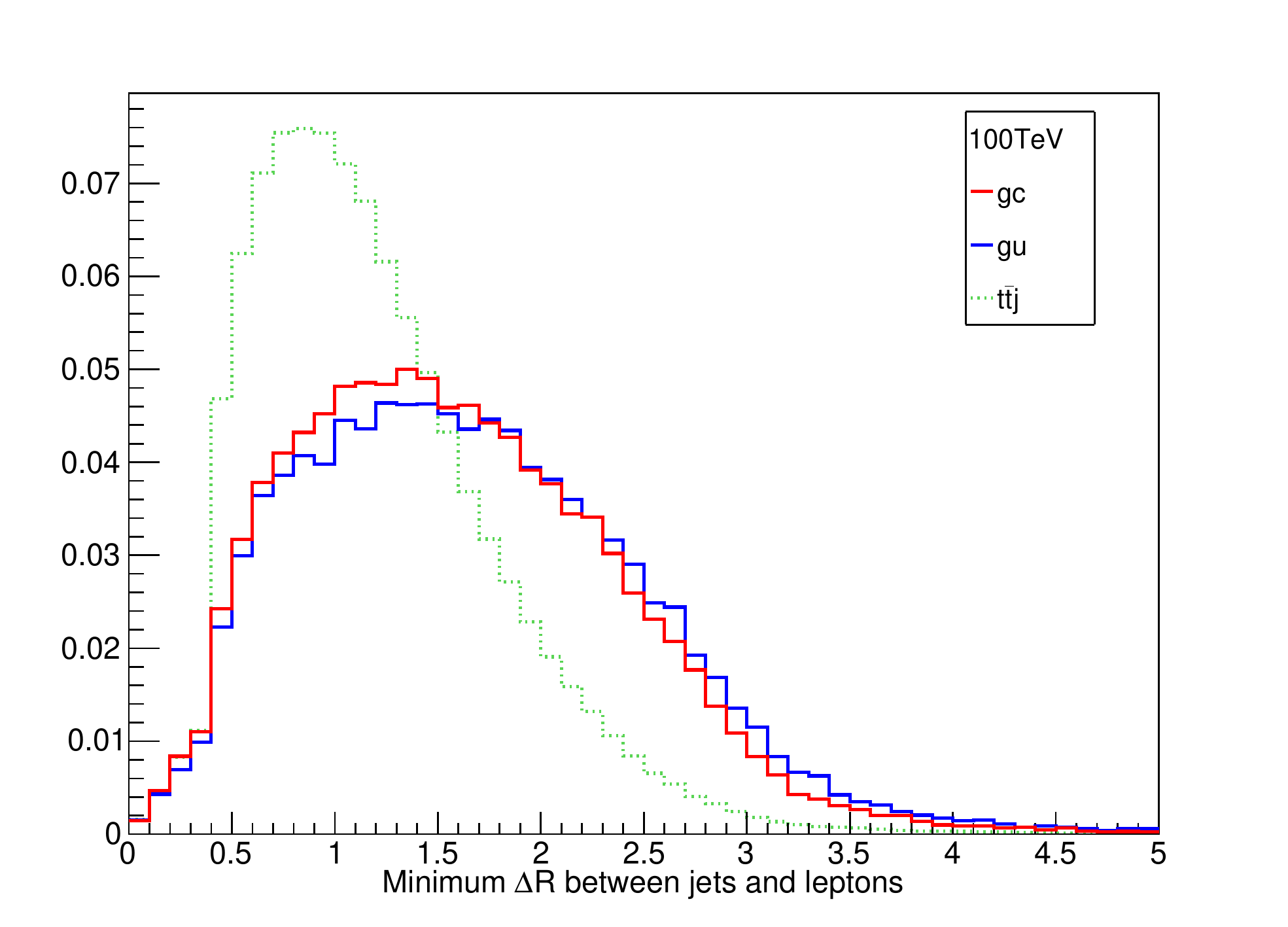}
\caption{The normalized distributions of $\Delta R_{\ell\ell}$ and $\Delta R_{\ell j}$ at 14 TeV and 100 TeV $pp$ colliders. }%
\label{fg:deltaR}%
\end{figure*}

The opening angle between the two charged leptons tends to be small due to the $V-A$ structure of the interaction between the leptons and $W$ boson~\cite{Dittmar:1996ss}. This can be used to separate the signal from the backgrounds where the charged leptons are likely to have a large opening angle. Furthermore, the Higgs and the top quark are boosted strongly if the collision energy is high. In the situation, the three jets are also close to each other and the leptons from Higgs should be moving more closely. In the center-of-mass frame, the top quark and Higgs boson are moving back to back, therefore the angular distance between the leptons and the jets would be far. In Fig.~\ref{fg:deltaR}, we show the distributions for $\Delta R_{\ell\ell}$ and $\Delta R_{\ell j}$. According to that, we employ the following cuts which are denoted as Cut-III,
\begin{align}
\Delta R_{\ell\ell}\leq 1.4,\quad \Delta R_{\ell j}\geq 1.8.\label{eq:wcutIII}
\end{align}

\begin{table}[h!]
\centering\begin{tabular}{|c|cccc|c|}
\hline
~ & Parton level & Cut-I & Cut-II & Cut-III &  $\epsilon_{III}$\\
\hline
$gc$& 72.37fb & 4.652fb & 0.7024fb & 200.7ab&	28.6\% \\
$gu$& 504.4fb & 25.62fb & 3.306fb & 1241ab	&	37.6\% \\
\hline
$t\bar{t}j$ & 56.23pb & 2866fb & 43.97fb &  6148ab& 14.0\%\\
wwjjj& 2.427pb & 14.24fb & 0.08803fb & 9.266ab& 10.6\%\\
zjjj& 687.5pb & 992.6fb & 9.499fb & -&-\\
\hline
\end{tabular}
\caption{The cross-sections of the signals and backgrounds with cuts on 14TeV $pp$ collider. The $gc$ ($gu$) denote the signal processes $gc\to tH$ and $g\bar{c}\to\bar{t}H$ ($gu\to tH$ and $g\bar{u}\to\bar{t}H$). The acceptance of the Cut-III $\epsilon_{III}$ is listed in the last column. The "-" means there is no event left after employing the cut, which requires more events to be generated for the analysis.}
\label{tb:ww14}
\end{table}
\begin{table}[h!]
\centering\begin{tabular}{|c|cccc|c|}
\hline
~ & Parton level & Cut-I & Cut-II & Cut-III & $\epsilon_{III}$ \\
\hline
gc & 2.711pb & 0.1554pb & 22.16fb & 8.839fb & 39.9\%\\
gu & 9.115pb & 0.2354pb & 36.34fb & 16.65fb &	45.0\%\\
\hline
$t\bar{t}j$ & 626.8pb & 14.89pb & 127.9fb & 13.79fb &10.8\% \\
\hline
\end{tabular}
\caption{The cross-sections of the signals and the main background with cuts on 100TeV hadron collider.The convention of the symbols is the same as that in Table.~\ref{tb:ww14}.}\label{tb:ww100}
\end{table}

The cross-sections for signals and backgrounds after the cuts are given in Table.~\ref{tb:ww14} for 14 TeV and Table.~\ref{tb:ww100} for 100 TeV $pp$ colliders. Only the dominated background $t\bar{t}j$ is represented for 100TeV collisions. All the events are generated at the tree level. And to avoid the soft-collinear divergence, we employ some loose parton-level cuts on the final quarks and leptons in Madgraph. These cuts are 
\begin{align}
p^i_T\geq 10\gev,\quad \Delta R_{ij}\geq 0.2.
\end{align}
where $i,j$ denote all the leptons and quarks in the final states.
The "-" means there is event left after employing the cuts. To obtain the cross-section, more events should be generated. Indeed we generate 290k events for the background $Zjjj$. Since there is no event left with the cuts, we can estimate a upper limit for the cross-section as 2.375fb with the event number $\leq1$. The acceptance for Cut-III is defined as the ratio between the cross-sections with and without the Cut-III. From that we can see that the cuts of the $\Delta R$s work better on 100 TeV than that on LHC. That can be understood as more Higgs bosons tends to be boosted at higher energy $pp$ collisions. 
\section{$h\to ZZ^*\to 4\ell^\pm$}
In this section, we make an analysis of the four charged leptons final states of the Higgs boson. The branching ratio of the Higgs decay $H\to ZZ^*\to 4\ell$ is 2 orders smaller than that of $H\to WW^*\to 2\ell$. This makes that such a channel is not promising on LHC. The final states are characterized by four charged leptons, one b-jet and two light jets. Such final states can be from $ZZ+X$ production in SM. However, the b-tag will suppress this background strongly. Here the background we consider are $pp\to ZZ+b\bar{b}+X$ and $pp\to ZV^*+b\bar{b}+X$ where $V=\gamma,Z$. We denote the total background as $Zllbbj$. Similar to the analysis in the above section, we need $N_\ell\geq 4$, $N_j\geq 3$ and $N_b=1$ and the same requirement as Eq.~\eqref{eq:WW_I}. These cuts are denoted as Cut-I.

\begin{figure*}[htb]%
\centering
\includegraphics[width=0.245\textwidth,trim={25 10 50 40},clip]{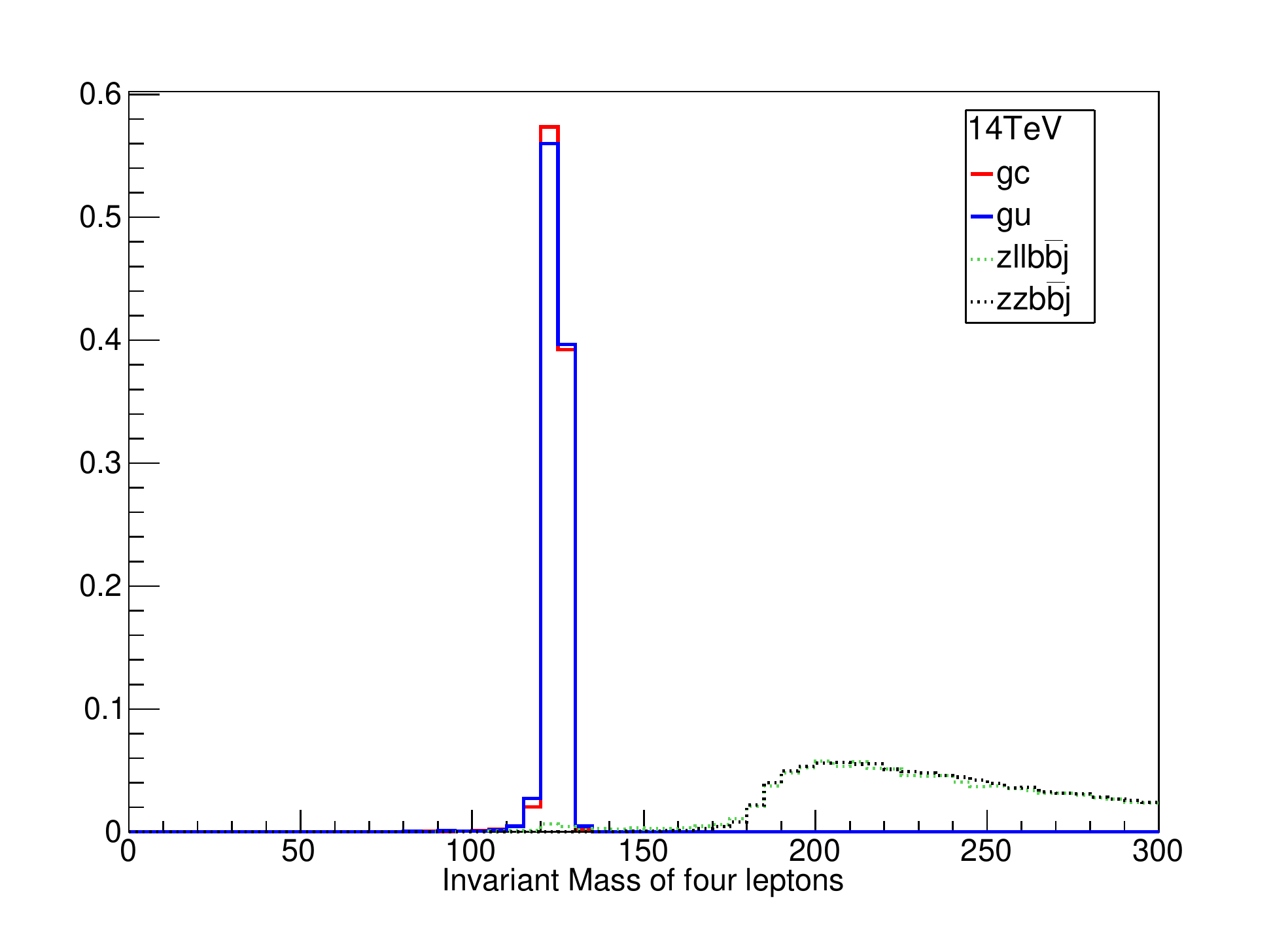}%
\includegraphics[width=0.245\textwidth,trim={25 10 50 40},clip]{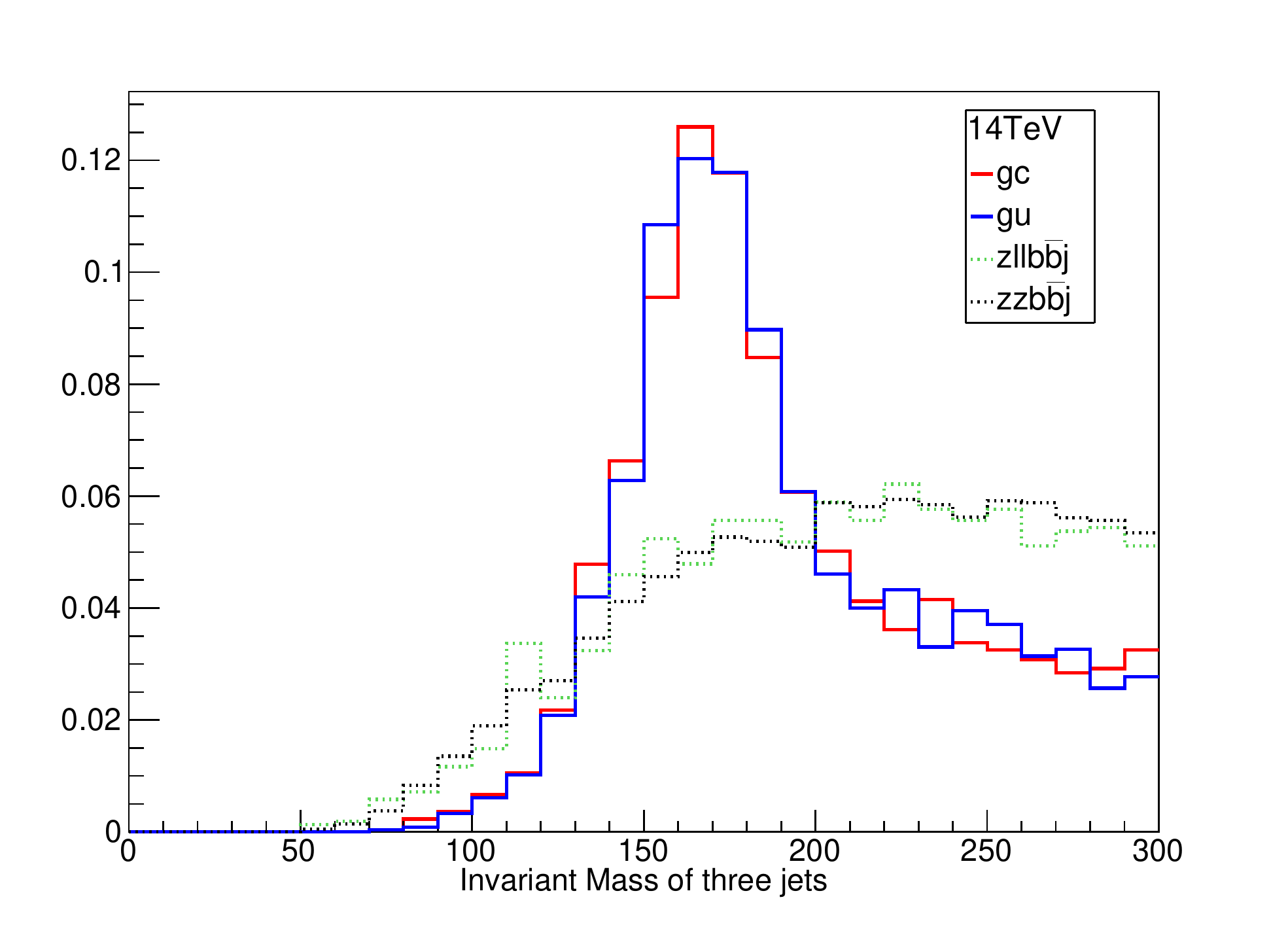}%
\includegraphics[width=0.245\textwidth,trim={25 10 50 40},clip]{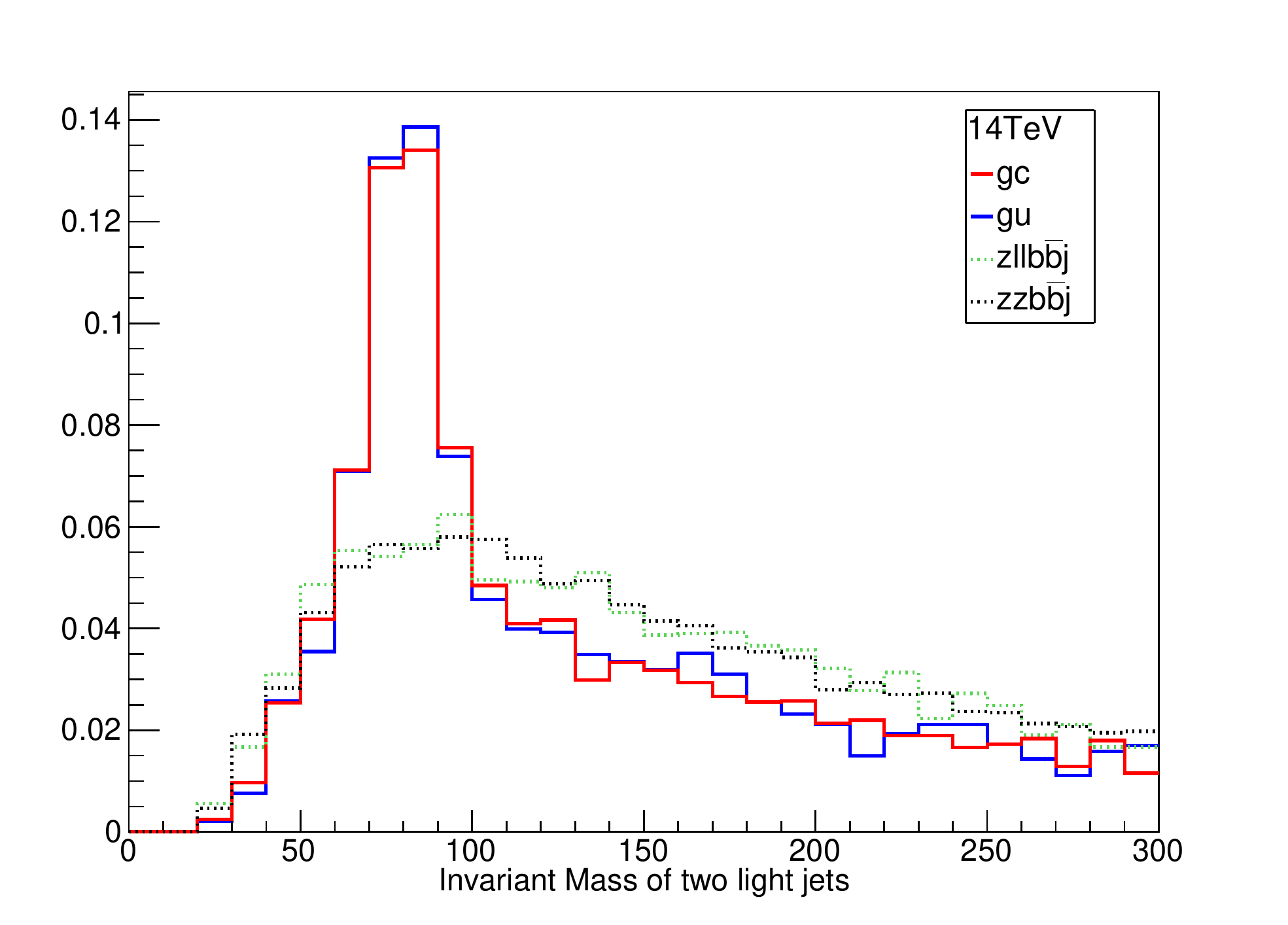}%
\includegraphics[width=0.245\textwidth,trim={25 10 50 40},clip]{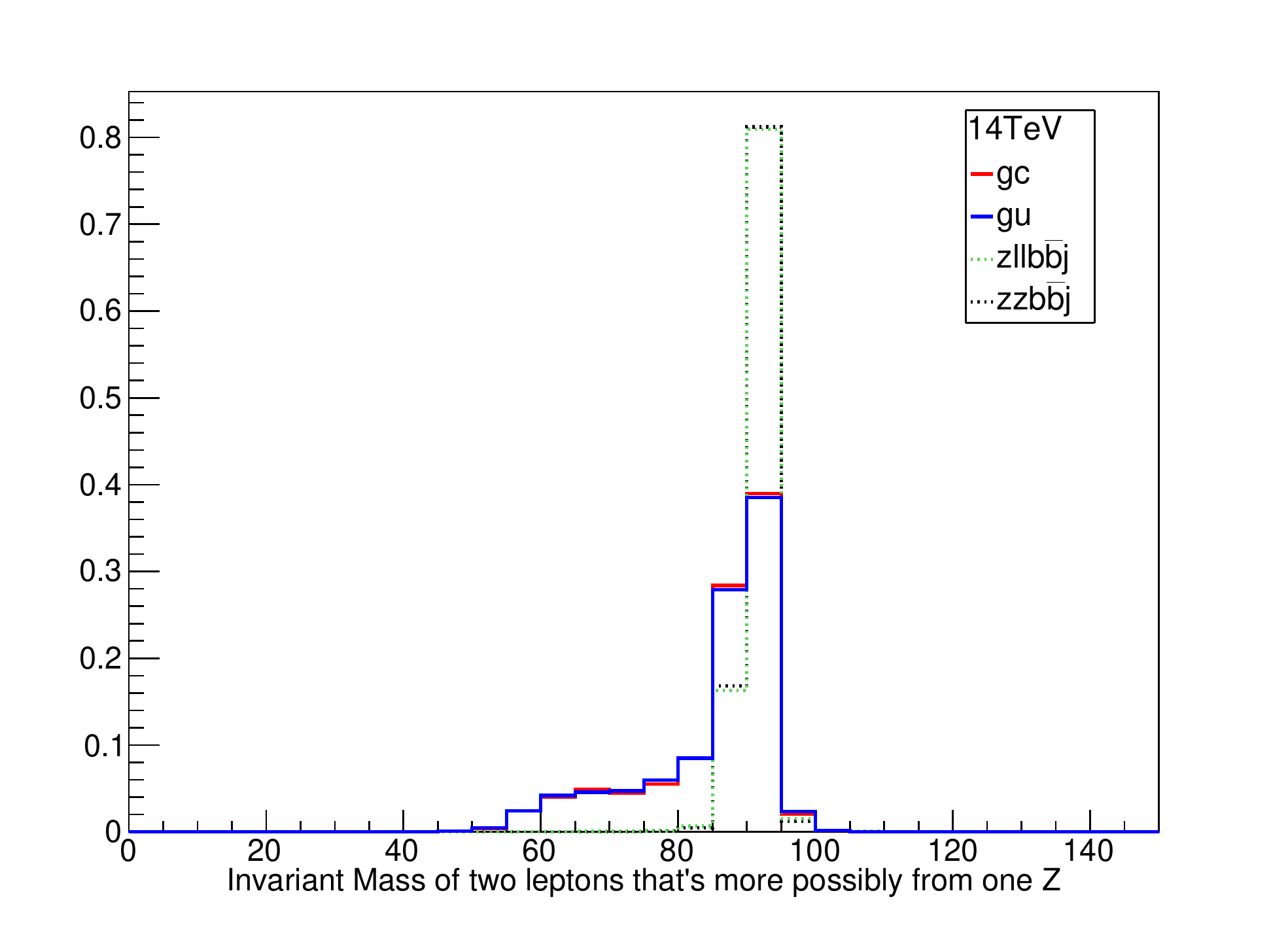}\\
\includegraphics[width=0.245\textwidth,trim={25 10 50 40},clip]{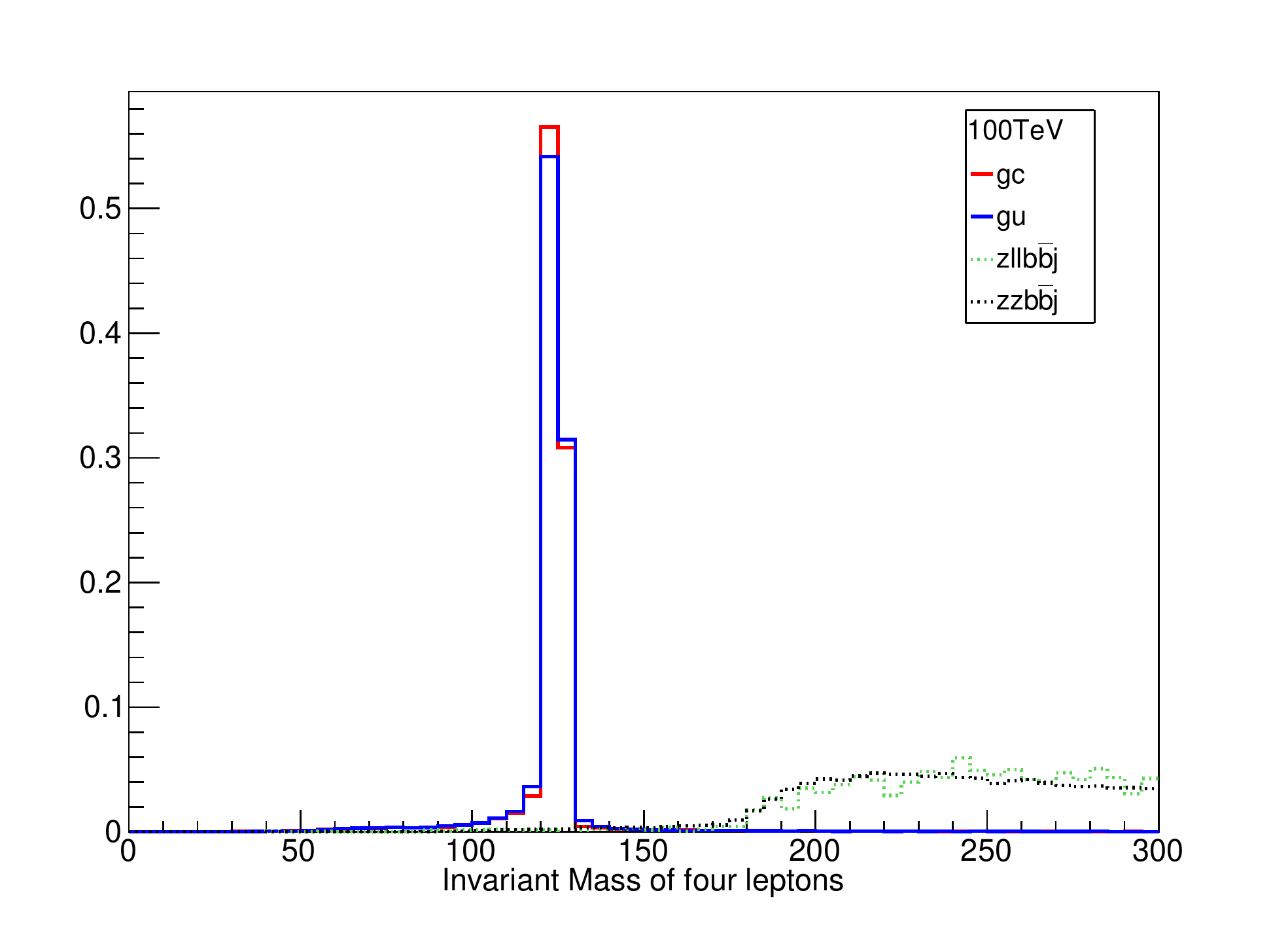}%
\includegraphics[width=0.245\textwidth,trim={25 10 50 40},clip]{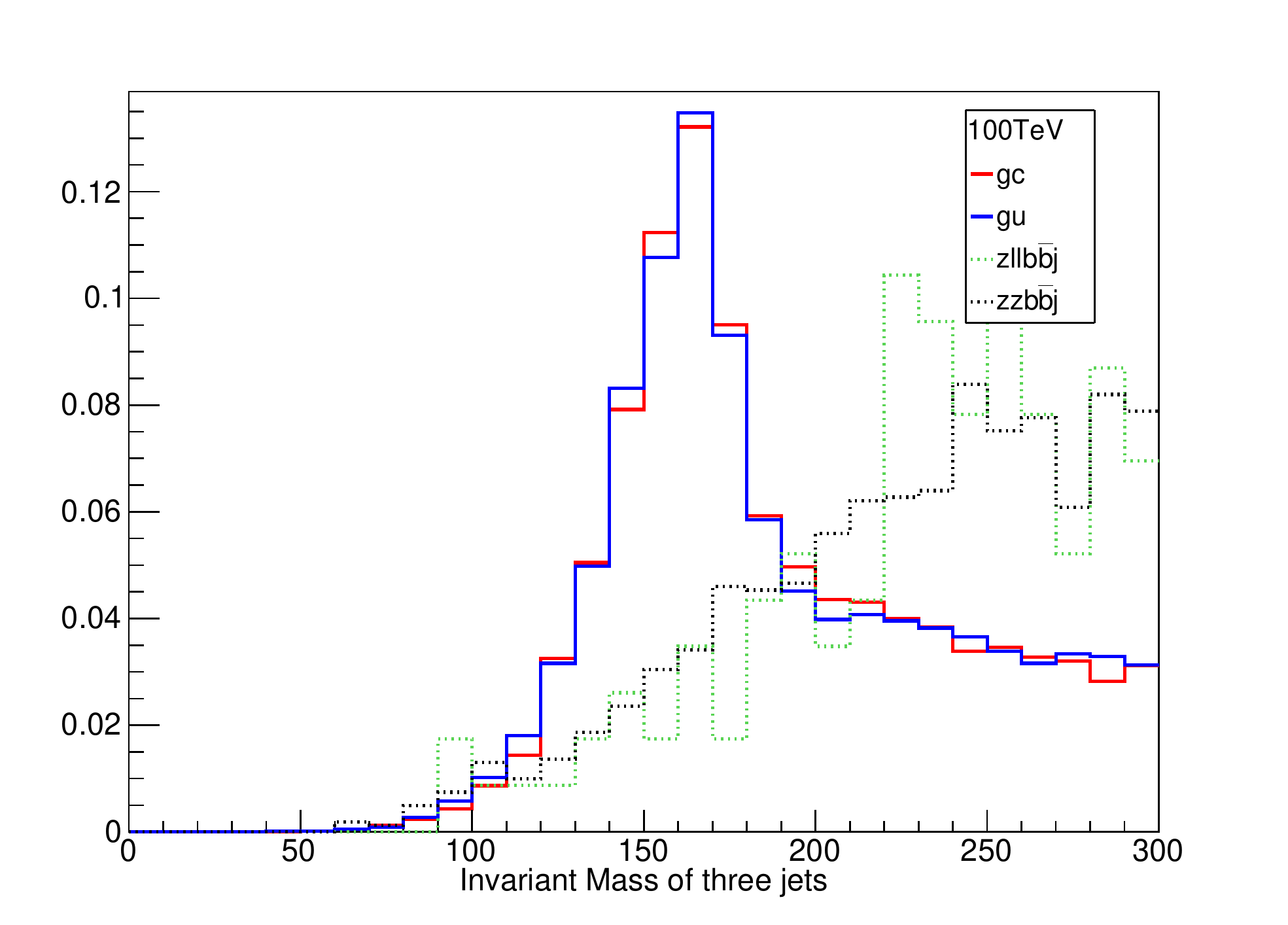}%
\includegraphics[width=0.245\textwidth,trim={25 10 50 40},clip]{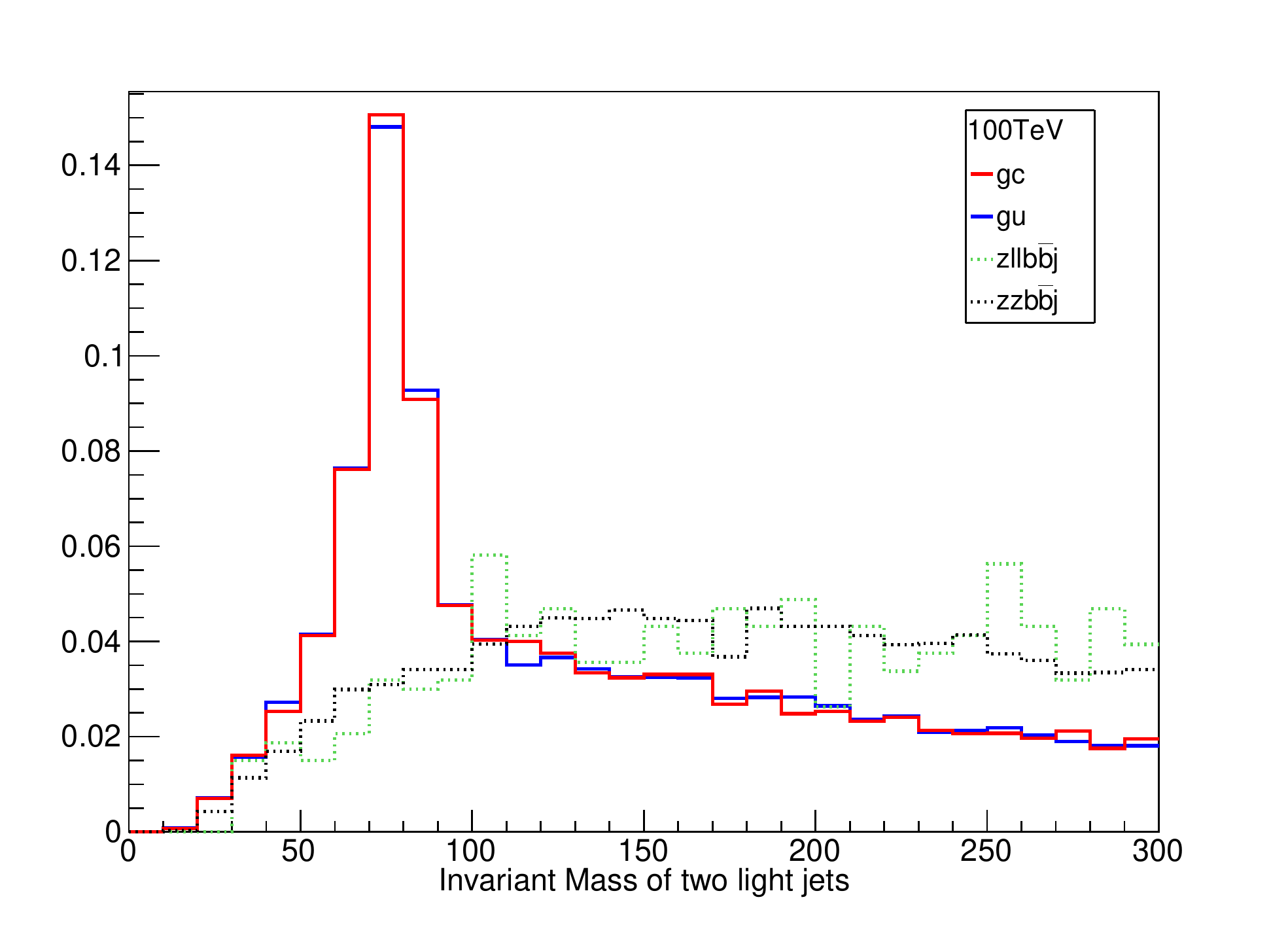}%
\includegraphics[width=0.245\textwidth,trim={25 10 50 40},clip]{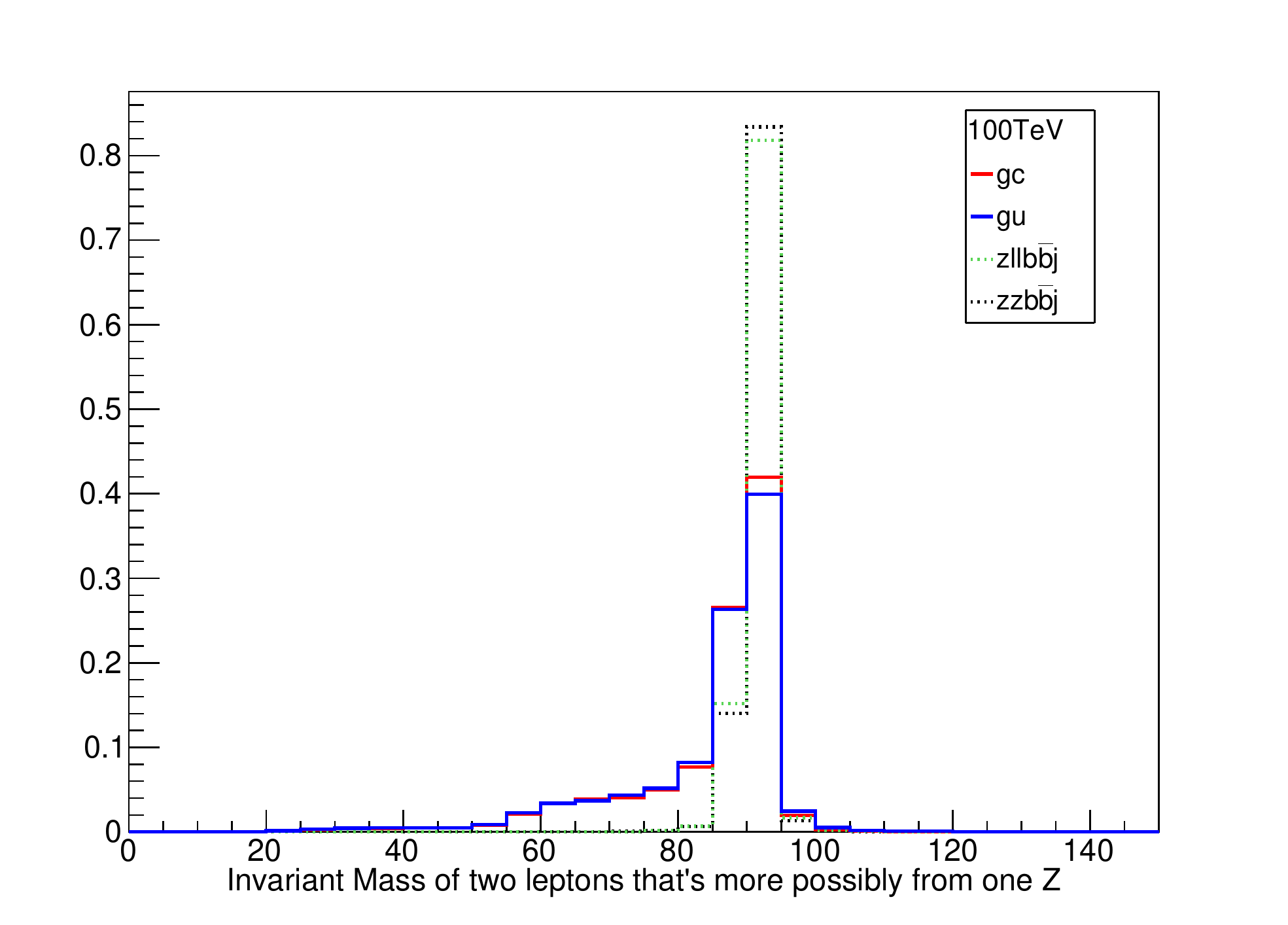}%
\caption{The normalized distributions for signal and background at 14TeV and 100TeV hadron colliders.}%
\label{fg:dist4ell}%
\end{figure*}

Moreover, we need the cut on the jets invariant mass $m_{3j}$ and the leptons invariant mass $m_{4l}$ to reconstruct the top quark and the Higgs boson. The distributions for the signals and backgrounds are shown in Fig.~\ref{fg:dist4ell}. 
Similar to the previous analysis, we choose a cut for the invariant mass of the two light $M_{jj}$ near $M_W$ and a cut for the invariant mass of the three jets $m_{bjj}$ near the top quark mass. There are two pairs of charged leptons in the signal, and one of them are from an on-shell $Z-$boson. We select the pair of charged leptons with same flavor whose invariant mass is most close to $M_Z$ to be the leptons from the on-shell $Z-$boson.
\begin{align}
 m^r_{z}=\{m_{\ell^+\ell^-}\vert \mathrm{min}(\vert m_{\ell^+\ell^-}-M_Z\vert) \}.
\end{align}
We collect the cuts as follows and named as Cut-II,
\begin{subequations}
\begin{align}\label{eq:zztop_higgs}
130\gev\leq m_{3j}\leq200\gev.\\
50\gev\leq m_{jj}\leq100\gev.\\
110\gev\leq m_{4\ell}\leq140\gev.\\
80\gev\leq m^r_{z}\leq100\gev. 
\end{align}
\end{subequations}
\begin{table}[h!]
\centering
\begin{tabular}{|c|cccc|}
\hline
~ & Parton level & Cut-I & Cut-II & Cut-III  \\
\hline
\multicolumn{5}{|c|}{14TeV}\\
\hline
gc & 1.554fb & 2.694ab & 0.5284ab & 0.1036ab  \\
gu & 10.80fb & 13.03ab & 2.303ab & 0.5758ab\\
\hline
$Z\ell\ell b\bar{b}j$ & 21.30ab & 5.705zb & 129.7yb & $\sim$0.3007yb\\
\hline
%
\multicolumn{5}{|c|}{100TeV}\\
\hline
gc & 58.12fb & 111.2ab & 23.63ab & 3.100ab  \\
gu & 195.4fb & 216.3ab & 49.51ab & 10.42ab\\
\hline
z$\ell\ell b\bar{b}$j & 48.1fb & 509.0ab & $\sim$208.3zb & $\sim$42.66yb\\
\hline
\end{tabular}
\caption{The cross-sections of the signals and backgrounds after employing the cuts. For efficiency, we employ $100\gev\leq m_{4\ell}\leq150\gev$ cut at parton level in Madgraph in addition to Eq.~\eqref{eq:zzparton}. The backgrounds are estimated roughly as explained in the text.}\label{tb:zz}
\end{table}

The four lepton in the background $ZZ+jets$ are from two on-shell $Z$ boson, therefore their invariant mass is greater than $2 M_Z$ as shown in Fig.~\ref{fg:dist4ell}. The Cut-II can veto most events of the process $ZZ+jets$.  Finally, we employ Cut-III which is the same as Eq.~\eqref{eq:wcutIII}. The cross-sections of the signals and backgrounds are shown in Table.~\ref{tb:zz}. One can see that the cross-sections of the signals with cuts seem too small to be detected on LHC. However, this channel could be detectable on 100TeV $pp$ colliders with high luminosity.
It should be clarified that the cross-sections for the backgrounds here are estimated very roughly. We generated 328476 events for LHC and 38258 events for 100TeV colliders with some parton-level cuts which are given as follows
\begin{align}
p^i_T\geq 10\gev,\quad\vert\eta ^i\vert\leq 5,\quad \Delta R_{ij}\geq 0.25.
\label{eq:zzparton}
\end{align}
And there is no event left with Cut I+II. To estimate the background with all the cuts, firstly we obtain the acceptance rates of three cuts separately and then assume the final acceptance rate as the product of the three cut acceptance rates.
\section{Results and discussion}
\begin{figure*}[htb]
\centering
\includegraphics[width=0.3\textwidth]{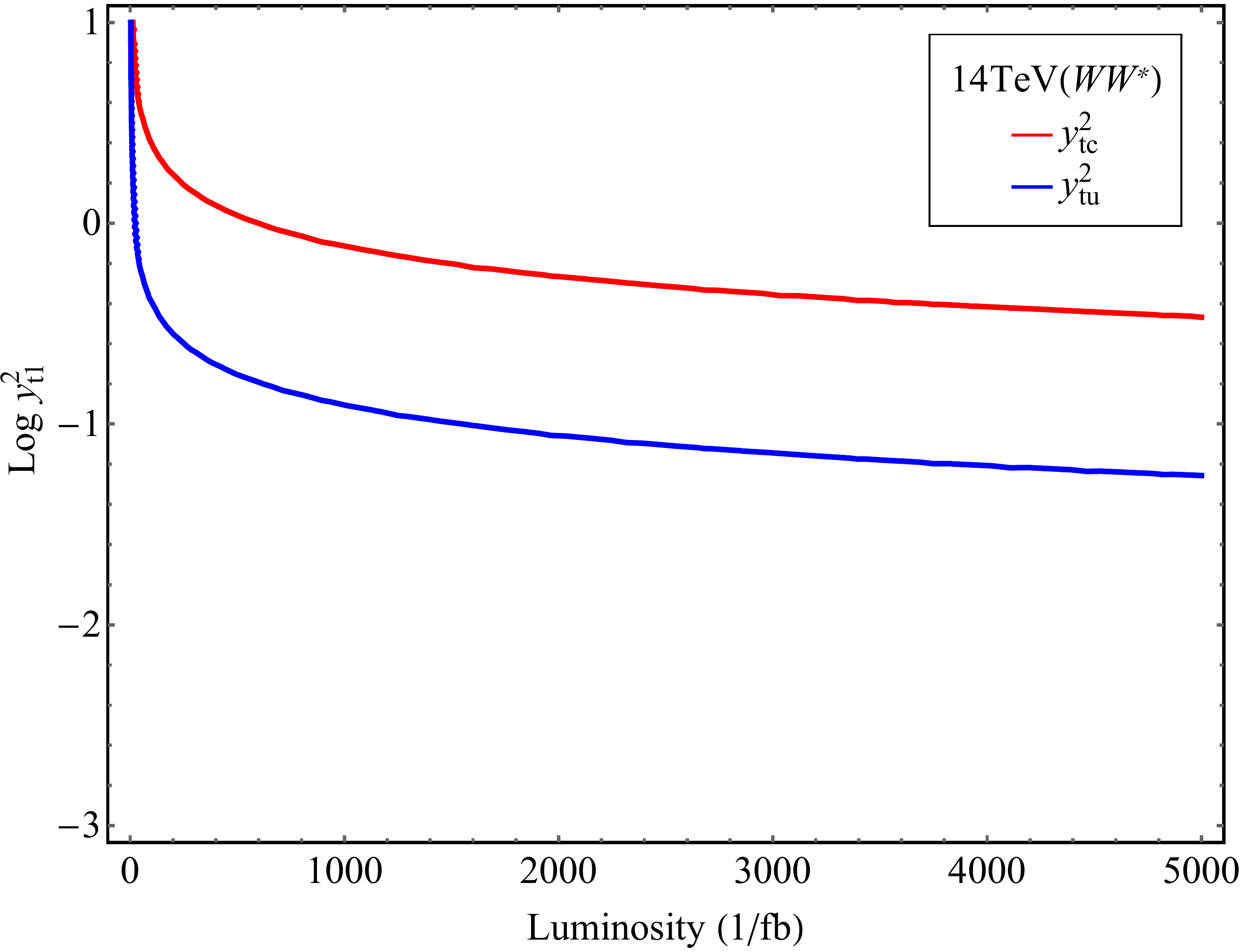}
\includegraphics[width=0.3\textwidth]{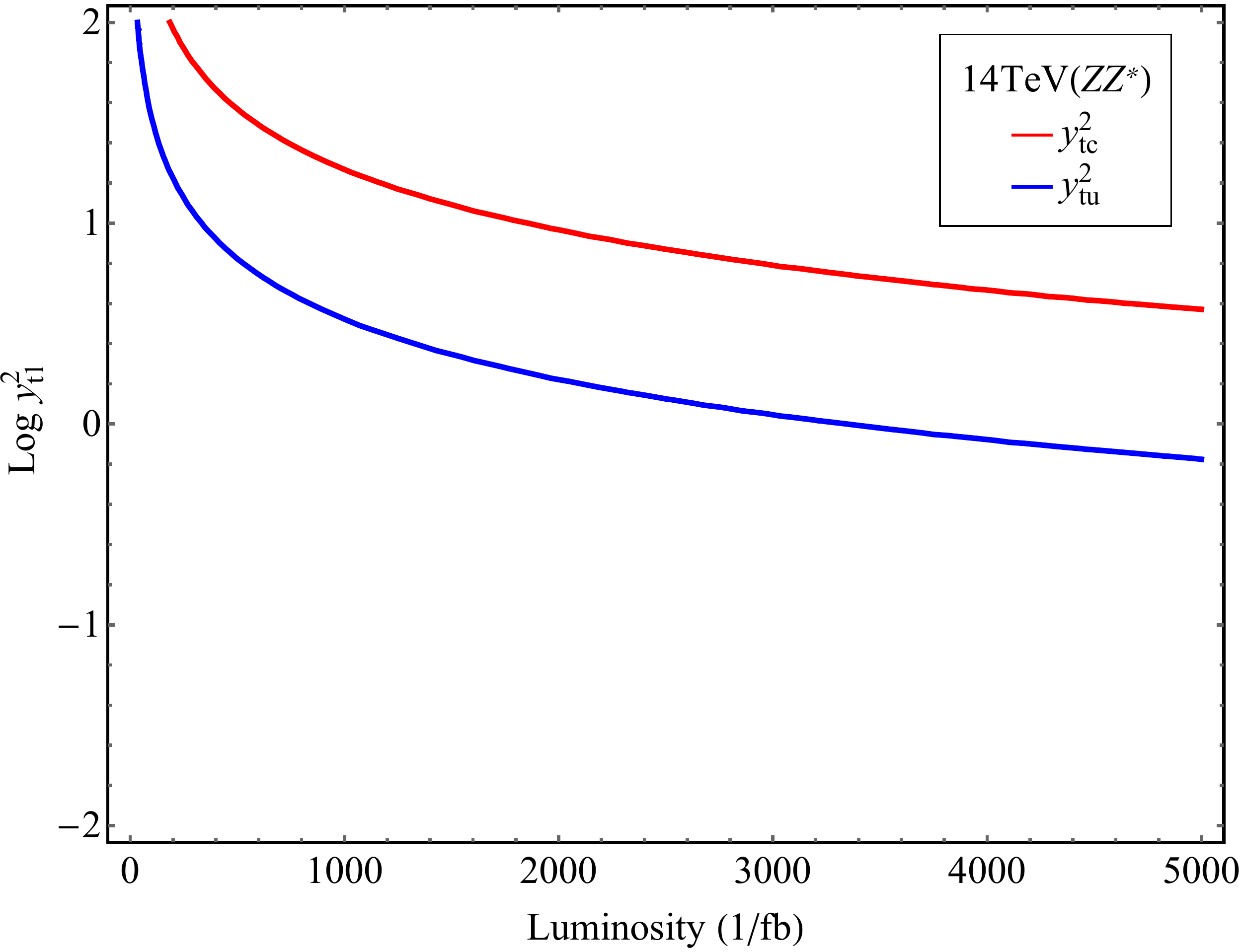}
\includegraphics[width=0.3\textwidth]{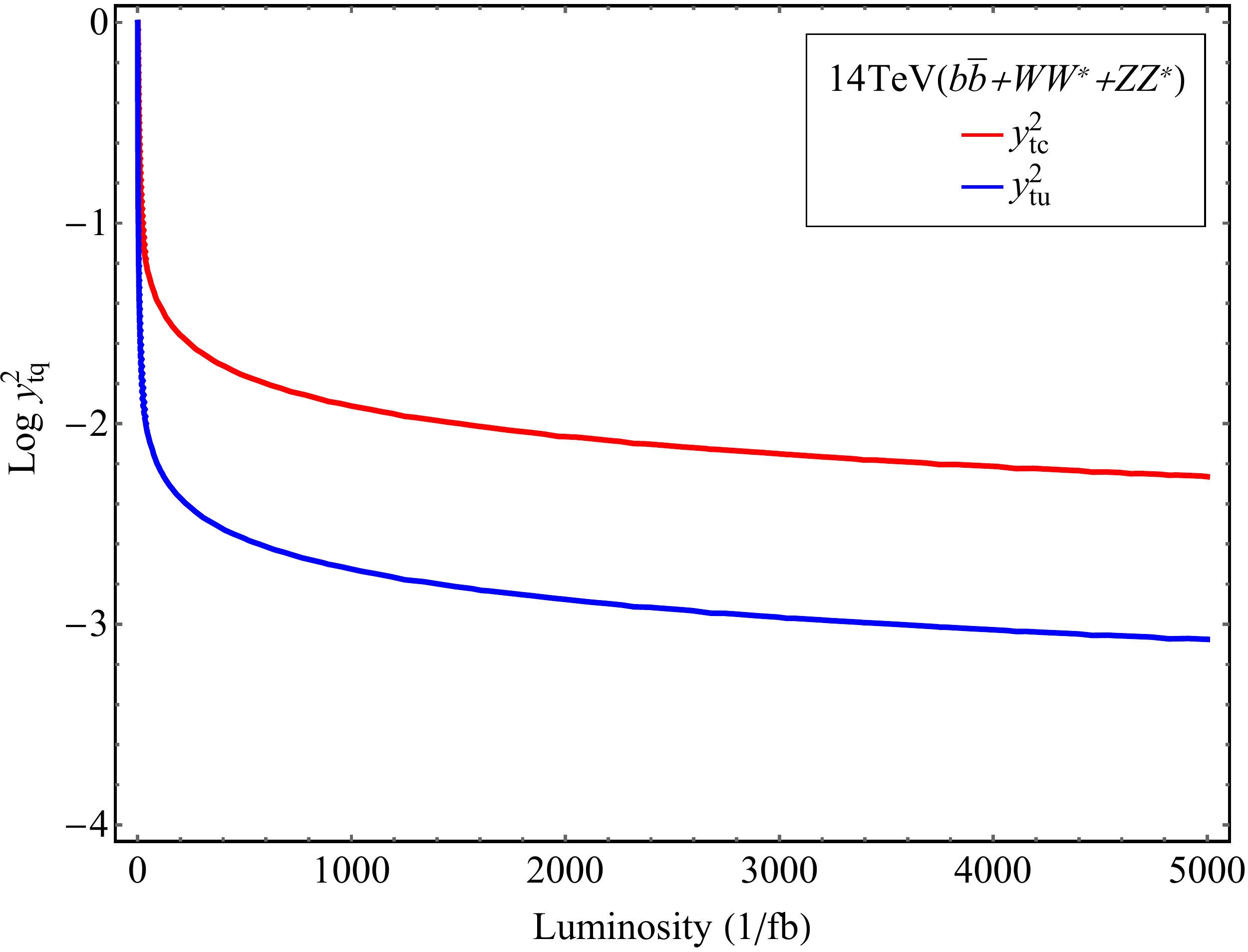}
\caption{The upper limits of $y_{tu}$ and $y_{tc}$ depend on the luminosity at LHC.}\label{fg:sgf14}
\end{figure*}

\begin{figure*}[htb]
\centering
\includegraphics[width=0.3\textwidth]{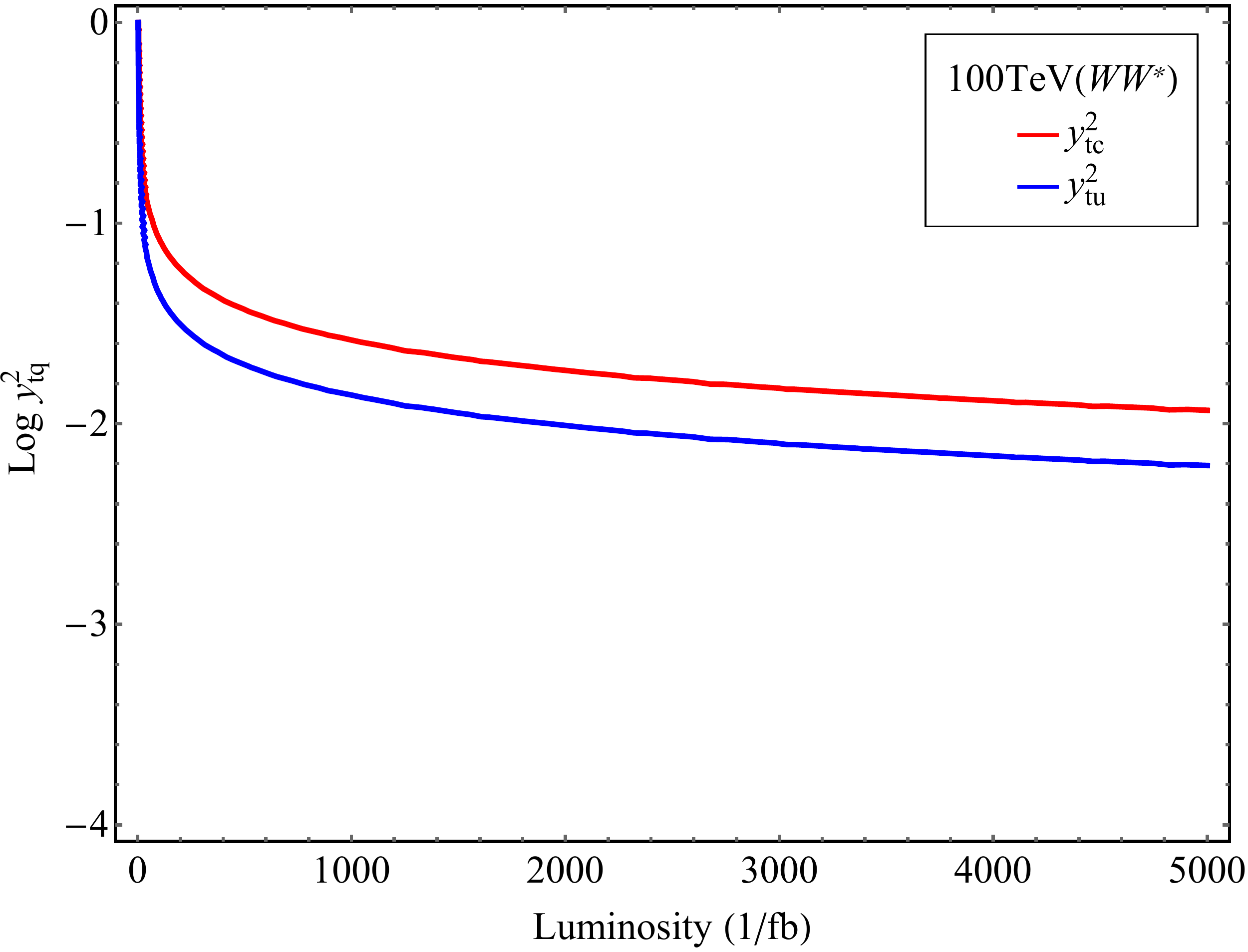}
\includegraphics[width=0.3\textwidth]{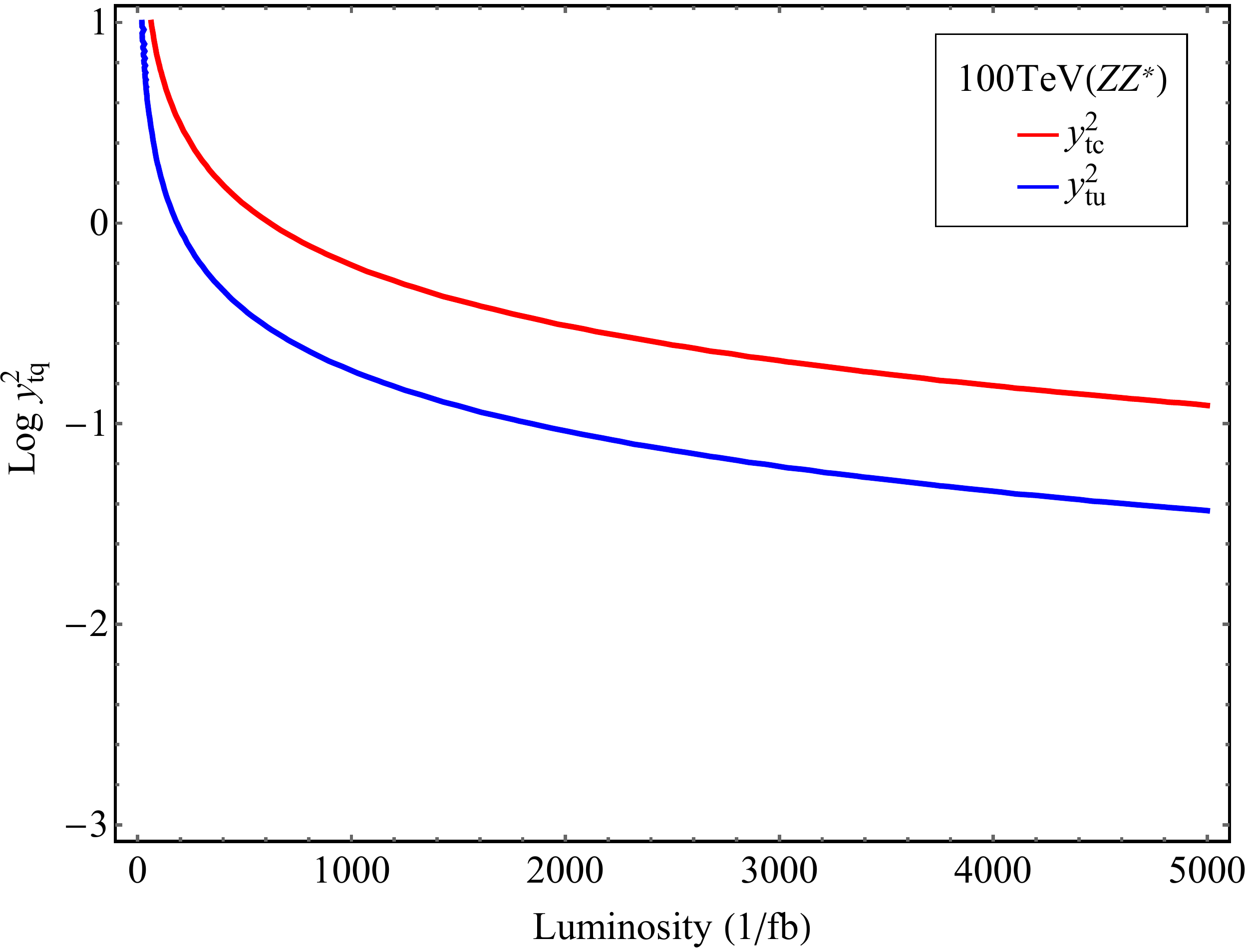}
\includegraphics[width=0.3\textwidth]{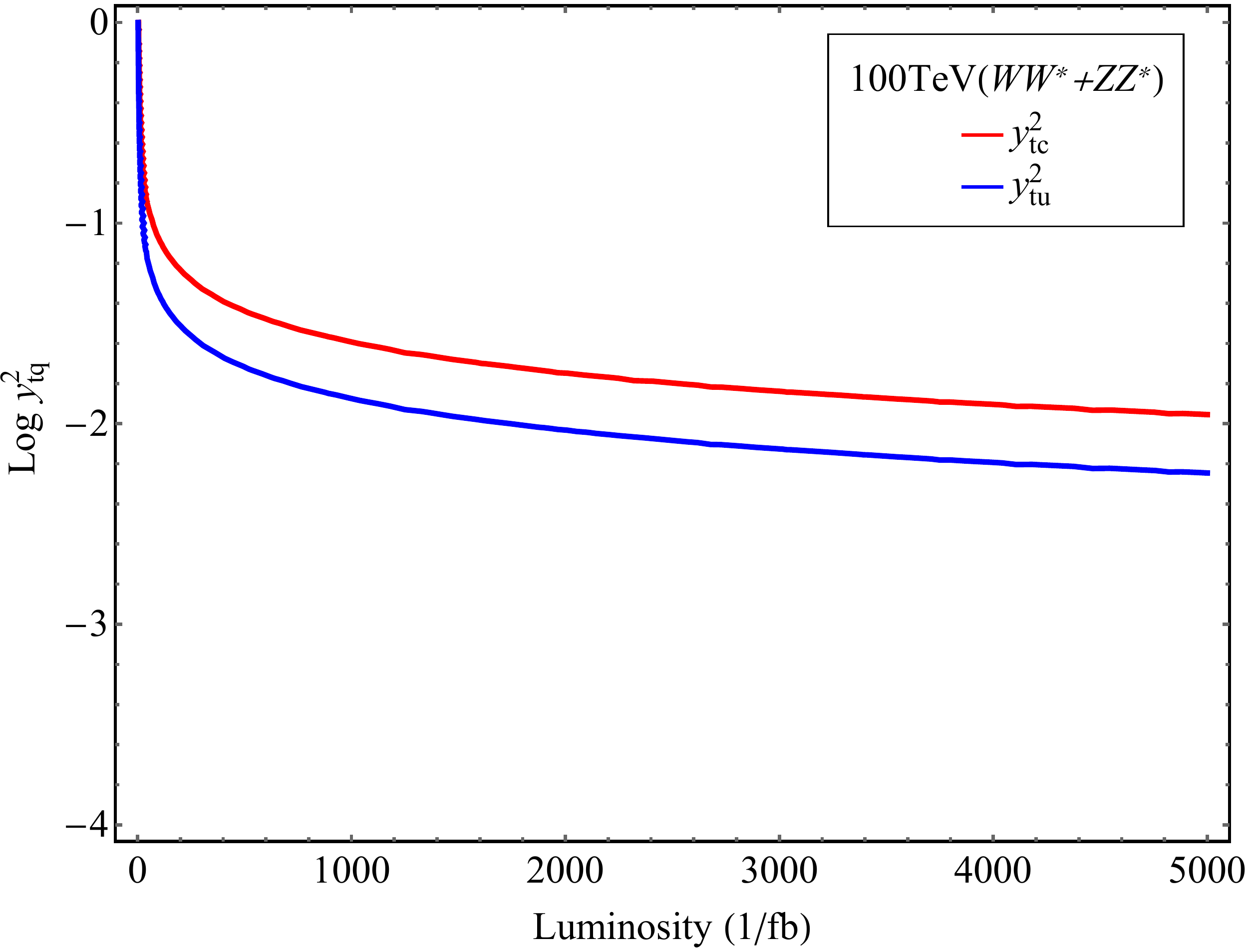}
\caption{The upper limits of $y_{tu}$ and $y_{tc}$ depend on the luminosity at 100TeV $pp$ collider.}\label{fg:sgf100}
\end{figure*}

From the previous analysis, the $WW^*$ channel is a promising channel to detect the FCNH couplings. Now we try to illustrate the detective potential of LHC and the hadron collider in the future. The likelihood function is defined as the Poisson probability~\cite{Cowan:2010js}
\begin{align}
L(n, n_b+n_s)=\frac{(n_b+n_s)^n}{n!} e^{-n_b-n_s}\label{eq:lkh}
\end{align}
where the $n$, $n_b$ and $n_s$ denote the observed data, SM background and signal events. In doing so we neglect the systematic uncertainties of the signal and the background estimation. To get the expected upper limit, we can replace $n$ with $n_b$.
The significance of the signal events relative to the background events is defined as
\begin{align}
\sigma_s(y_{tq})=\sqrt{-2\ln\frac{L(n_b,n_b+n_s)}{L(n_b,n_b)}}\simeq\frac{n_s}{\sqrt{n_s+n_b}} .
\end{align}
The values of $y_{tq}$ corresponding to $\sigma_s\leq2(3)$ can be regarded as the excluded region at 95\%(99\%) confidence level. 
As shown in Table.~\ref{tb:ww14}, Table.~\ref{tb:ww100} and Table.~\ref{tb:zz}, the $WW^*$ channel is more guaranteed than the $ZZ^*$ channel, and we can combine the $WW^*$ and $ZZ^*$ channels with the product of the likelihoods defined in Eq.~\eqref{eq:lkh}.The plots of the upper limits of the $y_{tu}$ and $y_{tc}$ depend on the luminosity of the hadron collider are given in Fig.~\ref{fg:sgf14} and Fig.~\ref{fg:sgf100}.

Combing the $WW^*$ and $ZZ^*$ channels, the upper limits of the FCNH couplings on $pp$ collider with high luminosity $\mathcal{L}=3000/fb$ at 95\% C.L. can be obtained as
\begin{align}
\text{14TeV:~}&y^2_{tu}\leq 6.96\times 10^{-2},\quad y^2_{tc}\leq  0.429 ,\\
\text{100TeV:~}&y^2_{tu}\leq  8.18\times 10^{-3},\quad y^2_{tc}\leq  0.0154.
\end{align}
We can see that this channel at LHC is not as good as the $H\to b\bar{b}$ channel studied in Ref.~\cite{Bao:2019hor}. Since then we combine all the three channels of $H\to b\bar{b}$, $WW^*$ and $ZZ^*$ and obtain the upper limits on LHC for $\mathcal{L}=3000/fb$ as,
\begin{align}
y^2_{tu}\leq 1.08\times 10^{-3},\quad y^2_{tc}\leq 7.08\times 10^{-3},
\end{align}
These constraints can be represented as the upper limits of the top decay,
\begin{align}
\mathrm{Br}(t\to H c)\leq& 4.09\times 10^{-3},\\
\mathrm{Br}(t\to H u)\leq& 6.06\times 10^{-4}.
\end{align}
According to the study in Ref.~\cite{ATLAS:2016qxw}, the constraints on the FCNC decays of the top quark could be reached to $Br(t\to H u)\leq 1.2 \times 10^{-4}$ and $Br(t\to H c)\leq 1.0\times 10^{-4}$ at 95\% C.L. on the LHC with $\mathcal{L}=3000/fb$. From that, we can see that the process $pp\to t/\bar{t}H$ seems not completable with the direct search for the decay $t\to H q$ on LHC. However, the couplings $y_{tu}$ and $y_{tc}$ can not be well distinguished in the top decay on the hadron colliders. Once the FCNC decay mode $t\to H q$ is measured, the $pp\to t/\bar{t} H$ can be used to identify the interaction of $y_{tq}$.

\section{Summary}
In this work we studied the FCNC Yukawa couplings of the top quark at 14TeV and 100TeV hadron colliders through the process $pp\to t/\bar{t}H$. We analyze the leptonic decay channels of the Higgs boson. We find that the special region where the Higgs and top are boosted works well to distinguish the signal and backgrounds. Such region can be characterized by the angular distances $\Delta R\leq 1.4$ of the leptons from the Higgs boson and $\Delta R>1.8$ between the final jet from top quark and leptons from the Higgs boson. And these cuts work better on 100TeV hadron collider. We also checked the possibility to search this process at the hadron colliders. We found that the $WW^*$ channel is a promising channel on both 14TeV and 100TeV colliders while the cross-section of $ZZ^*$ channel seems a little small. We combined different channels and obtained the expected upper limits of the couplings. These limits can be presented as $Br(t\to H c)\leq 4.09\times 10^{-3}$ and $Br(t\to H u)\leq 6.06\times 10^{-4}$. We can see that the process $pp\to t/\bar{t} H$ is not as good as $t\to H q$ in setting the limits of $y_{tq}$. However, this process can be used to distinguish the $y_{tq}$ couplings, if the FCNC decay of the top quark $t\to Hq$ is observed.
\begin{acknowledgments}
This work is supported in part by the Natural Science Foundation of Shandong Province under grant No. ZR2020MA094.
\end{acknowledgments}

\end{document}